\documentclass[letterpaper]{JHEP3}
\usepackage{nicefrac}
\usepackage{amsmath,epsfig}
\usepackage{amssymb,amsfonts}

\usepackage{graphics}
\usepackage{epsfig}
\usepackage{amsfonts}

\newcommand{\nn}{\nonumber}

\newcommand{\Tr}{\mbox{Tr}}

\def\d{\partial}

\def\s{\sigma}

\def\a{\alpha}
\def\b{\beta}
\def\e{\epsilon}

\def\h{\eta}

\def\half{{\frac12}}

\def\IC{\relax\hbox{$\inbar\kern-.3em{\rm C}$}}

\def\IC{{\bf C}}

\def\bea{\begin{eqnarray}}
\def\eea{\end{eqnarray}}
\def\be{\begin{equation}}
\def\ee{\end{equation}}
\def\ba{\begin{align}}
\def\ea{\end{align}}
\def\bse{\begin{subequations}}
\def\ese{\end{subequations}}
\def\1F1{{}_1\!F_1}
\def\2F0{{}_2\!F_0}

\def\ni{\noindent}

\def\nn{\nonumber}

\def\a{\alpha}

\def\h3{$\textrm{H}_3^+$}
\def\d{{\partial}}

\def\IC{{\mathbb C}}

\def\Tr{{\rm Tr}}

\def\lbldef#1#2{\expandafter\gdef\csname #1\endcsname {#2}}

\def\href#1#2{#2}

\newcommand{\beq}{\begin{equation}}
\newcommand{\eeq}{\end{equation}}
\newcommand{\ber}{\begin{eqnarray}}
\newcommand{\eer}{\end{eqnarray}}

\def\be{\begin{eqnarray}}
\def\ee{\end{eqnarray}}

\def\({\left(}
\def\){\right)}
\def\[{\left[}
\def\]{\right]}
\def\<{\langle}
\def\>{\rangle}
\def\d{\partial}

\def\gg{\mathsf g }

\title{Extremal Correlators  and Hurwitz Numbers \\ in  Symmetric Product Orbifolds}
\preprint{Brown-HET-1582 \\ YITP-SB-09-12}
\author{
Ari Pakman\footnote{Email: ari$\_$pakman@brown.edu}$^{~1}$, Leonardo Rastelli\footnote{Email: leonardo.rastelli@stonybrook.edu}$^{~2}$, and
Shlomo S. Razamat\footnote{Email: razamat@max2.physics.sunysb.edu}$^{~2}$
\\ \\ \\
\it $^1$ Department of Physics,\\ Brown University,\\
Providence, RI 02912, USA
\\
\\
\it $^2$ C.N. Yang Institute for Theoretical Physics,\\
\it Stony Brook University, \\
\it Stony Brook, NY 11794-3840, USA}

\abstract{

We study correlation functions of single-cycle chiral operators in ${\rm Sym}^N \,T^4$,
the symmetric product orbifold of $N$ supersymmetric four-tori.
Correlators of twist operators
 are evaluated on  covering surfaces, generally of different genera, where fields are single-valued.
We compute some simple four-point functions  and study how the sum over inequivalent branched covering maps splits under OPEs.
We then discuss extremal $n$-point correlators, {\it i.e.} correlators of~$n\!-1$ chiral and one anti-chiral operators.
They obey simple recursion relations involving numbers obtained from counting branched covering maps with particular properties.
In most cases we are able to solve explicitly the recursion relations. Remarkably, extremal correlators turn out to be equal to
 Hurwitz numbers.
}


\begin{document}
%\maketitle \setcounter{tocdepth}{2}
%\tableofcontents

\section{Introduction}

In this paper we apply the technology developed in a companion article \cite{PRR1} to
study  correlation functions of single-cycle twist operators in ${\rm Sym}^N \, T^4$,
 the symmetric product orbifold of $N$ supersymmetric four-tori.  We focus
 on {\it extremal} correlators, which by definition are correlators
 of $n\!-\!1$ chiral and one anti-chiral operators, where the notion of chirality
 is that of a $(2,2)$ subalgebra of the full $(4,4)$ supersymmetry.

An important motivation for this work is the holographic duality between the symmetric product of $T^4$ and
type  IIB string theory on $AdS_3 \times S^3 \times T^4$~\cite{Maldacena:1997re}.\footnote{See
\cite{Aharony:1999ti, Dijkgraaf:2000vr,David:2002wn, Martinec} for reviews.}
The early checks of this duality included comparison of the moduli spaces \cite{Dijkgraaf:1998gf, Larsen:1999uk}
and the spectra of both theories~\cite{Maldacena:1998bw, deBoer:1998ip, Kutasov:1998zh, Argurio:2000tb}.
Recently much progress was made in comparing
correlation functions. The structure constants of single-cycle operators in the
chiral ring of the symmetric product
were computed early on in~\cite{Jevicki:1998bm} and, for a subset of these operators
 they were extended in~\cite{Lunin:2000yv, Lunin:2001pw} to the full 1/2 BPS $SU(2)$ multiplet. These three-point functions were exactly reproduced
in the string theory/supergravity
dual~\cite{Gaberdiel:2007vu, Dabholkar:2007ey, Pakman:2007hn, Taylor:2007hs} (see also~\cite{Giribet:2007wp,  Giribet:2008yt, Cardona:2009hk}),
which also predicts some correlators not yet computed in the symmetric product~\cite{Pakman:2007hn}.

This agreement between bulk and boundary correlators was at first
surprising because the computations on the string and CFT sides are performed at very different points in the moduli space~\cite{Dijkgraaf:1998gf,Larsen:1999uk}.  It can  be explained by a non-renormalization theorem proved in~\cite{deBoer:2008ss}.
The non-renormalization theorem  also holds for extremal correlators (as first conjectured in~\cite{Taylor:2007hs}), so
we expect that they can be successfully compared  on both sides of the $AdS_3/CFT_2$ correspondence as well.

Extremal correlators play a special role in the AdS/CFT correspondence.
Not only are they not renormalized:
they offer a unique window into the bulk/boundary dictionary, as we review at the end of this introduction.
The main purpose of this paper is to compute extremal correlators of (anti)chiral twist fields on the CFT side of the duality. Explicit
computations in symmetric product orbifolds can be notoriously difficult   (see {\it e.g.} \cite{Arutyunov:1997gt,Lunin:2000yv, Lunin:2001pw}).
As usual in orbifold theories, determining a correlator involves finding branched covering maps,
 with branching points fixed by the position and type of the different twist fields.
The non-abelian character of the permutation group  implies that starting from three-point functions
 several branched coverings, generally of different genera,  contribute to a given correlator.

In general finding the relevant covering maps is a challenging task
to pursue analytically. When the covering surface is a sphere, which provides the leading contribution for large $N$,
the covering maps are quotients of polynomials whose coefficients depend on  the
length of the cycles in the correlator  (see \cite{PRR1} for a thorough discussion).

However, extremal correlators are special.
First, their genus zero covering maps are among the simplest: either a polynomial or a polynomial divided by a monomial.
Moreover, in the former case the {\it only} contribution to the correlators comes from the sphere.
Second,
extremal correlators  have trivial spacetime dependence,
and only the overall constant needs to be determined. One of our main results is that finding this constant does not require detailed knowledge of the branched covering
 maps. What is needed are the total number of maps and  the numbers of maps contributing to the OPEs between two operators.
As we will show, these numbers enter in simple recursion relations that determine all the extremal correlators of single-cycle operators.

The problem of computing extremal correlators  thus reduces to counting branched covering maps
with given branching structure (and other additional restrictions).
These enumeration questions have a long history~\cite{hur} and are generally referred to as the { ``Hurwitz problem''}.
The Hurwitz problem has an equivalent reformulation in terms of enumerating sets of elements of $S_N$
that multiply to the identity. Symmetric product orbifolds provide a very intuitive picture of
this equivalence \cite{PRR1}.

We are able to compute most extremal correlators of single-cycle twist operators,
by finding the solutions to the relevant enumeration problems when the covering surface is a sphere.
The final expressions have a remarkable property: after a certain rescaling of the (anti)chiral operators, all  correlators
are equal to the relevant Hurwitz number, {\it i.e.} to the total number of maps from the covering surface!
We conjecture that this result holds also for extremal correlators of multi-cycle twist operators.
We emphasize that the simple final answer arises non-trivially after combining partial results.
As always,   miracles of this kind call for a deeper explanation. We suspect that
a topologically twisted version of the symmetric orbifold CFT will be the natural framework
to understand why extremal correlators compute Hurwitz numbers.
We note in this respect that a topologically twisted version of the {\it worldsheet} theory with $AdS_3 \times S^3$ target \cite{Rastelli:2005ph}
computes correlators of spacetime chiral primary operators. It would be interesting
to see if the methods of \cite{Rastelli:2005ph} can be used to simplify the calculation of extremal correlators on the string theory side.

We conclude this introduction with some general remarks about the role of extremal correlators in the AdS/CFT correspondence.\footnote{It is
also interesting
 to note that in the context of Gopakumar's approach to string duals of free field theories~\cite{Gopakumar:2005fx} extremal correlators
 play a (technically) preferred role due to the relative simplicity of obtaining them~\cite{Aharony:2006th,David:2006qc,David:2008iz}.}
It is worth recalling
the situation in the $AdS_5/CFT_4$ instance of the duality. Extremal correlators in ${\cal N} = 4$ SYM
(and in the dual IIB string theory on $AdS_5 \times S^5$)
are also believed to obey a non-renormalization theorem. Evidence
for their protection came first from  their calculation on the supergravity side \cite{D'Hoker:1999ea},
which gives a result proportional to the free-field expression (whereas a non-trivial dependence
on the cross ratios would be a priori expected), and it was confirmed by explicit perturbative
calculations on the field theory side \cite{Bianchi:1999ie}, as well as by formal arguments using ${\cal N}=2$ harmonic
superspace~\cite{Eden:1999kw, Eden:2000gg}.  See also \cite{Erdmenger:1999pz, D'Hoker:2000dm, D'Hoker:2000vb} for related work.
What makes extremal correlators especially interesting is not just their protection,
but the fact  that they are uniquely sensitive to the color structure of  operators. In ${\cal N} = 4$ SYM chiral primary operators
of charge $k$ (under a $U(1) \subset SU(4)$ subgroup the $R$-symmetry)  are linear combinations
of the form
\begin{equation} \label{admixture}
\Tr Z^k + \frac{1}{N} \sum_{\ell}  \, a_\ell \, \Tr Z^{k-\ell} \, \Tr Z^\ell + \frac{1}{N^2}  \sum_{\ell_1, \ell_2} \, a_{  \ell_1 \ell_2 }\Tr Z^{k-\ell_1-\ell_2} \, \Tr Z^{\ell_1}  \, \Tr Z^{\ell_2} + \dots\, .
\end{equation}
If the coefficients $a_{\ell_1 \dots \ell_m}$ are taken to be independent of $N$, each term in the sum
has the same large $N$ scaling (since each trace contributes a factor of $N$). However
when computing large $N$ correlation functions, the usual factorization arguments imply that
{\it generically} only the single-trace piece of the operator contributes. Extremal correlators are the exception to this rule:
the multi-trace admixtures contribute at the same order \cite{D'Hoker:1999ea}.  The dual statement on the gravity side
is that extremal correlators are uniquely sensitive to boundary terms, which can instead
be neglected in the non-extremal cases \cite{D'Hoker:1999ea}.

Finding the precise dictionary between bulk states and boundary operators requires knowing which precise admixture
of single and multi-trace operators map to single-particle states in the bulk. Extremal correlators
 offer a window to find this precise dictionary.  It was found in \cite{D'Hoker:1999ea} by a careful evaluation of supergravity boundary terms
in extremal 3-point functions in $AdS_5 \times S^5$
that at leading order in $N$
one can match  {\it single}-trace  operators of ${\cal N} = 4$ SYM to bulk {\it single}-particle states:
 in other terms $a_\ell = O(1/N)$. By contrast in little string theory the holographic dictionary requires mapping
 single string states with admixtures of the form (\ref{admixture}) with $a_{\ell_1 \dots \ell_m}$ of order one \cite{Aharony:2004xn}.

 In the symmetric product orbifold, twist operators are classified by conjugacy classes
 of the symmetric group $S_N$, {\it i.e.} by their cycle structure -- with the single-cycle
 operators playing a somewhat similar role as the single-trace operators in a gauge theory.
 In $AdS_3/CFT_2$, the agreement between the string theory
  and the field theory calculations of three-point functions
 works by postulating the naive correspondence between single-strings and single-cycles \cite{Gaberdiel:2007vu, Dabholkar:2007ey, Pakman:2007hn}
 (in other tems the analogue of $a_\ell$ must be assumed to be zero at large $N$ to find agreement).
 Most {\it supergravity} extremal three-point correlators also match assuming the naive
 dictionary \cite{Taylor:2007hs}. The matching of some doubly exceptional supergravity 3-point correlators (extremal 3-point correlators containing
 states that saturate the Breitenlohner-Freedman bound) appears instead to require an order-one admixture of single and  double-cycles \cite{Taylor:2007hs}.\footnote{Alternatively, this may just indicate
 a subtlety in the naive supergravity calculation, which was performed by analytic continuation away from extremality.
 A careful analysis of supergravity boundary terms is necessary to
  confirm the validity of the analytic continuation procedure, indeed we would expect boundary terms  to be especially subtle for states sitting exactly at the BF bound.}
 We hope that the extremal $n$-point correlators computed in this paper will be useful in sharpening the bulk-to-boundary dictionary.

The paper is organized as follows. In Section 2 we review the basic observables and interactions in the single-cycle sector of the chiral ring.
In Section 3,
 we compute some non-extremal four-point functions and extract useful
 lessons about the structure of the OPE  in symmetric product orbifolds.
In Section 4 we compute extremal correlators of single-cycle operators, discover that they compute Hurwitz numbers
and present a general conjecture  for arbitrary extremal correlators.
Two appendices collect technical results used in the body of the paper.

\section{Single-cycle states in the chiral ring}
The symmetric product orbifold ${\rm Sym}^N \,T^4$ is obtained by considering $N$ copies of $T^4$ and identifying the coordinates under
the action of the permutation group $S_N$. Each copy of $T^4$ has bosonic coordinates $X_{I}^i$ with real fermionic partners $\chi_I^i$,
where $i=1,2,3,4$ and $I=1,\ldots,N$.\footnote{ For more details on symmetric product orbifolds we refer the reader to~\cite{PRR1} and references therein. For more details on the chiral ring  of the symmetric product of $T^4$ see~\cite{Jevicki:1998bm}.}
The basic observables of a symmetric product orbifold  are the
twist fields~${\s}_{[g]}$, labeled by a conjugacy class $[g]$ of the permutation group.
Conjugacy classes with one non-trivial cycle of length $n$ will be denoted by~$[n]$.
Clearly the  OPE of two single-cycle  twist-fields gives both single-cycle and multi-cycle operators,
a fact that will play a prominent role
in the calculation of extremal correlators (see  Section~\ref{extremal}).

``Gauge-invariant'' twist fields~$\s_{[g]}$ can be constructed from ``gauge-non-invariant'' ones, $\s_g$, associated
to a group element $g\in S_N$ and not to a conjugacy class.
Single- and double-cycle  group elements have the form
\be
g_{single} &=& (12 \ldots n)\,,
\\
g_{double}  &=& (12 \dots n_1)(n_1\!+\!1 \ldots n_1\!+\!n_2)\,,
\
\ee
and similarly for higher-cycle operators. This representation indicates on which of the~$N$ elements the cyclical permutations act, and in which order.
We refer to  the  set of values of~$I$ on which~$g$ acts  as ``colors'' of an element~$g$.
The operator $\s_g(z,\bar z)$ is defined as a ``defect'' imposing
 the following monodromies on the
different copies of the fields:
\be
X^i_I(e^{2\pi i}\,z)\s_g(0)= X^i_{g(I)}( z)\s_g(0) \,,
\ee
and similarly   for the fermionic fields.
Single cycle ``gauge-invariant''  operators  are obtained by  averaging over the group orbit,
\be
\label{gaugeinvariant}
\s_{[n]} \equiv
%{\mathcal A}_{[g]}(N)\,
\frac{1}{\sqrt{n\,N!(N-n)!}} \sum_{h\in S(N)}\s_{h^{-1}(12 \ldots n)h} \, ,
\ee
where the pre-factor gives the normalization
\be
\langle  \s_{[m]}(0)  \s_{[n]}(z)\rangle =\frac{ \delta_{mn}}{{|z|^{2 \Delta_n} } }\,.
\ee
There is a similar definition for double-cycle operators.
The operators (\ref{gaugeinvariant}) have conformal dimension (see e.g.~\cite{Arutyunov:1997gt})
\be
\Delta_n = \frac{6}{24}\left(n-\frac{1}{n} \right),
\ee
where the $6$ in the numerator is the central charge of each copy of supersymmetric $T^4$.

The (anti)chiral operators are built by dressing the twist-fields~(\ref{gaugeinvariant}) with invariant contributions from the fermionic sector to
satisfy the (anti)chiral relationship $\Delta= \pm Q$, where $Q$ is the charge under the $U(1)$ of the ${\cal N}=2$ subalgebra.
There are three types of chiral operators: $O_n^{(0,0)}$, $O_n^{(a,\bar{a})} (a,\bar{a}=1,2)$ and $O_n^{(2,2)}$,
corresponding to $0,1$ and $2$-forms in~$T^4$, respectively, with $n$ being the length of the permutation cycle.
We will consider only operators whose holomorphic and antiholomorphic quantum numbers are equal,
but it is easy to extend our results to operators of mixed type, {\it e.g.} $O_n^{(0,2)}$.

More explicitly,  the four real holomorphic fermions of $T^4$ can be combined, in each copy $I$, into two complex fermions $\psi^1_I, \psi^2_I$, and bosonized as
\be
\psi_I^1&=& e^{i\phi_I^1}\,,
\\
\psi_I^2&=& e^{i\phi_I^2}\,, \qquad \qquad I=1,\ldots, N\,.
\ee
The $U(1)$ current of the ${\cal N}=2$ algebra is
\be
J&=& \frac{i}{2}  \sum_{I=1}^N \d \phi_I^1 + \d \phi_I^2\,.
\ee
We define first the gauge-non-invariant chiral operators,
\be
o_{(12\ldots n)}^{(0,0)} &=&  e^{ i \frac{n-1}{2n} \sum_{I=1}^{n} (\phi_I^1 +  \phi_I^2  + \bar{\phi}_I^1 + \bar{\phi}_I^2 )}   {\sigma}_{(12\ldots n)}\, ,
\label{gnop1}
\\
o_{(12\ldots n)}^{(a=1,\bar{a}=1)} &=&  e^{ i \frac{n+1}{2n} \sum_{I=1}^{n} (\phi_I^1 + \bar{\phi}_I^1 ) + i \frac{n-1}{2n} \sum_{I=1}^{n}(\phi_I^2  +\bar{\phi}_I^2)}  {\sigma}_{(12\ldots n)}\, ,
\\
o_{(12\ldots n)}^{(a=2,\bar{a}=2)} &=&  e^{ i \frac{n-1}{2n} \sum_{I=1}^{n} (\phi_I^1 + \bar{\phi}_I^1 ) + i \frac{n+1}{2n} \sum_{I=1}^{n}(\phi_I^2  + \bar{\phi}_I^2)} {\sigma}_{(12\ldots n)}\, ,
\\
o_{(12\ldots n)}^{(2,2)} &=&  e^{ i \frac{n+1}{2n} \sum_{I=1}^{n} (\phi_I^1 +  \phi_I^2  +\bar{\phi}_I^1 + \bar{\phi}_I^2)}  {\sigma}_{(12\ldots n)}\,.
\label{gnop4}
\ee
and the gauge invariant operators are obtained by summing over the group orbit as in (\ref{gaugeinvariant}),
\be
O_n^{(0,0)} &=&  \frac{1}{\sqrt{n\,N!(N-n)!}} \sum_{h\in S(N)}   o_{h^{-1} (12\ldots n)h}^{(0,0)}
\, ,
\label{op1}
\\
O_n^{(a,\bar{a})} &=&  \frac{1}{\sqrt{n\,N!(N-n)!}} \sum_{h\in S(N)}   o_{h^{-1} (12\ldots n)h}^{(a,\bar{a})}\, ,
\\
O_n^{(2,2)} &=&  \frac{1}{\sqrt{n\,N!(N-n)!}} \sum_{h\in S(N)}   o_{h^{-1} (12\ldots n)h}^{(2,2)} \,.
\label{op4}
\ee
The conformal dimensions  and charges are
\be
\Delta^{0}_n &=& Q^{0}_n = \frac{n -1 }{2}\, ,
\label{jepsn}
\\
\Delta^{a}_n &=& Q^{a}_n = \frac{n}{2}\, ,
\\
\Delta^{2}_n &=& Q^{2}_n = \frac{n +1 }{2}\, ,
\ee
and similarly for the antiholomorphic sector. The antichiral operators
$O_n^{(0,0) \dagger}, O_n^{(a,\bar{a}) \dagger}, O_n^{(2,2) \dagger}$
are obtained by reversing the sign in the exponents in (\ref{gnop1})-(\ref{gnop4}).
The theory has actually ${\cal N}=4$ supersymmetry, and the (anti)chiral  states are the (lowest)highest weights in an $SU(2)$ multiplet.

The fusion rules of the chiral ring are~\cite{Jevicki:1998bm}
\be
(0,0) \times (0,0) &\rightarrow& (0,0) + (2,2) \nn\, , \\
(0,0) \times (2,2) &\rightarrow& (2,2)\, ,
\label{fusionrules}
\\
(0,0) \times (a,a) &\rightarrow& (a,a) \nn\, , \\
(a,a) \times (a,a) &\rightarrow& (2,2) \nn\,.
\ee
These rules are easy to obtain by combining the composition law of the permutation group and the conservation of $U(1)$ charge.
To  leading order in $1/N$ the five structure constants  corresponding to the above OPEs are~\cite{Jevicki:1998bm}
\be
\langle O_{n_{3}}^{(0,0) \dagger} O_{n_{2}}^{(0,0)}  O_{n_{1}}^{(0,0)} \rangle
&=& \left( \frac{1}{N} \right)^{\frac{1}{2}} \frac{(n_3)^{3/2}}{( n_{2} n_1 )^{1/2}}\, ,
\label{nbos1}
\\
\langle O_{n_{3}}^{(2,2) \dagger} O_{n_{2}}^{(2,2)} O_{n_{1}}^{(0,0)}  \rangle
&=& \left( \frac{1}{N} \right)^{\frac{1}{2}} \frac{(n_{2})^{3/2}}{(n_3   n_1 )^{1/2}}\, ,
\label{nbos3}
\\
\langle O_{n_{3}}^{(b,\bar{b}) \dagger} O_{n_{2}}^{(a,\bar{a})}  O_{n_{1}}^{(0,0)} \rangle
&=& \delta^{a b}  \delta^{\bar{a} \bar{b}} \left( \frac{1}{N} \right)^{\frac{1}{2}} \frac{(n_3 n_{2})^{1/2}}{(n_{1})^{1/2}}\, ,
\label{nfer1}
\\
\langle O_{n_{3}}^{(2,2) \dagger} O_{n_{2}}^{(a,\bar{a})} O_{n_{1}}^{(b,\bar{b})}  \rangle
&=& \e^{ab} \e^{\bar{a}\bar{b}}  \left( \frac{1}{N} \right)^{\frac{1}{2}} \frac{(n_{2} n_{1})^{1/2}}{(n_3 )^{1/2}}\, ,
\label{nfer2}
\\
\langle O_{n_{3}}^{(2,2) \dagger} O_{n_{2}}^{(0,0)} O_{n_{1}}^{(0,0)} \rangle
&=& \left( \frac{1}{N} \right)^{\frac{1}{2}} \frac{1}{(n_3 n_{2}  n_1 )^{1/2}}\, .
\qquad \qquad
\label{nraro}
\ee
Conservation of $U(1)$ charge imposes the relation $n_3= n_1 + n_2 -1$ in all the cases, except in~(\ref{nraro}), where we have $n_3=n_1+n_2-3$.

Consider a correlator of gauge-invariant operators (ignoring the fermionic dressing and normalization factors) ,
\be\label{corr}
\langle\prod_{j=1}^{p} \s_{[n_j]} (z_j, \bar z_j) \rangle \sim
\sum_{h_j\in S(N)} \, \langle \prod_{j=1}^p \,  \s_{h_j (1 2 \ldots n_j) h_j^{-1}} (z_j, \bar z_j) \rangle \, .
\label{sumcorr}
\ee
A term in this expansion will be non-zero only if the product of its group elements {\it in a fixed order}\footnote{
It is convenient  to choose this order to coincide with the radial ordering of the operators in the correlator (see \cite{PRR1}
for details).}
 satisfies,
\be
(n_p) (n_{p-1}) \ldots (n_1)=1\, ,
\label{prodone}
\ee
where  $(n_j) \equiv h_j (1 2 \ldots n_j) h_j^{-1}$.

Some terms of the sum (\ref{sumcorr}) will be ``disconnected'', namely, the  $p$-point function splits into two or more independent factors with no common colors, such as
\be
 \langle \prod_{j=1}^q \,  \s_{h_jg_jh_j^{-1}} (z_j, \bar z_j) \rangle_{conn}
\langle \! \prod_{j=q\!+\!1}^p \,  \s_{h_jg_jh_j^{-1}} (z_j, \bar z_j) \rangle)_{conn}\, .
\label{discoloco}
\ee
Each term  in the expansion (\ref{sumcorr}) has a certain number of active  colors, which we denote by $c$.
For example, in a three-point function, the term
\be
\langle \s_{(4321)}  \s_{(34)} \s_{(123)} \rangle \, ,
\ee
has $c=4$.
It is convenient to organize the connected terms in the sum (\ref{sumcorr}) into groups of terms with fixed number of colors
\be
\langle\prod_{i=1}^{p} \s_{[n_i]} (z_i, \bar z_i) \rangle_{conn.} = \sum_{c} \, R_c(n_i)\, ,
\label{sumrc}
\ee
such that each $R_c$ is a sum of terms
\be
R_c(n_i) = F_c (n_i, N) \sum_{j=1}^{H_c} \langle \s_{n_p} \ldots \s_{n_1} \rangle_j\, ,
\label{rc}
\ee
with a fixed $c$.  $H_c$  is the number of distinct ways of satisfying (\ref{prodone}), up to a relabeling of colors reflected in the symmetry factors $F_c (n_i,N)$,
which can be computed exactly (see~\cite{PRR1} for details). These symmetry factors encode the dependence of the correlators on $N$.

The terms in (\ref{rc}) are computed, as usual in orbifolds, by going % from the sphere $z$
 to the covering surface(s) where operators are single-valued.
The
 genus $\gg$ of the  covering  surface is fixed by the Riemann-Hurwitz formula
\be\label{rh1}
\gg=\half \sum_{j=1}^p \left(n_j-1\right)-c+1 \,.
\ee
Thus we see that the sum over $c$ in (\ref{sumrc}) is equivalently a sum over the genera of the covering surfaces.
For every $\gg$ the symmetry factor scales at large $N$ as (see {\it e.g.} \cite{PRR1})
\be
F_c (n_i, N) \sim N^{1-\gg -\frac{p}{2}}\, ,
\label{largen}
\ee
which shows that the leading terms in (\ref{sumrc}) come from the sphere ($\gg=0$).

An interesting property of the four  structure constants (\ref{nbos1},\,\ref{nbos3},\,\ref{nfer1},\,\ref{nfer2})
of the chiral ring, which will hold for their generalization to $p$-point extremal correlators as well, is that only
genus zero covering surfaces contribute. Indeed, the Riemann-Hurwitz relation~(\ref{rh1}) and the relation $n_3=n_1+n_2-1$ give
\be
\gg=\half \sum_{j=1}^3 \left(n_j-1\right)-c+1 = n_3 -c\,.
\ee
But the number of colors $c$ is at least as big as the longest cycle, $c \geq n_3$, and therefore only the $c=n_3, \gg=0$
term in the genus expansion contributes. Therefore, using the proper~$F_c(n_i)$ factors, one
 can write expressions valid for finite $N$~\cite{Jevicki:1998bm},
\be
\langle O_{n_{3}}^{(0,0) \dagger} O_{n_{2}}^{(0,0)}  O_{n_{1}}^{(0,0)} \rangle
&=& F(n_1,n_2)  \frac{(n_3)^{3/2}}{( n_{2} n_1 )^{1/2}}\, ,
\label{bos1}
\\
\langle O_{n_{3}}^{(2,2) \dagger} O_{n_{2}}^{(2,2)} O_{n_{1}}^{(0,0)}  \rangle
&=& F(n_1,n_2)  \frac{(n_{2})^{3/2}}{(n_3   n_1 )^{1/2}}\, ,
\label{bos3}
\\
\langle O_{n_{3}}^{(b,\bar{b}) \dagger} O_{n_{2}}^{(a,\bar{a})}  O_{n_{1}}^{(0,0)} \rangle
&=& \delta^{a b}  \delta^{\bar{a} \bar{b}} F(n_1,n_2) \frac{(n_3 n_{2})^{1/2}}{(n_{1})^{1/2}}\, ,
\label{fer1}
\\
\langle O_{n_{3}}^{(2,2) \dagger} O_{n_{2}}^{(a,\bar{a})} O_{n_{1}}^{(b,\bar{b})}  \rangle
&=& \e^{a b}  \e^{\bar{a} \bar{b}} F(n_1,n_2) \frac{(n_{2} n_{1})^{1/2}}{(n_3 )^{1/2}}\, ,
\label{fer2}
\ee
where
\be
F(n_1,n_2) = \left[ \frac{(N-n_1)! (N-n_2)!}{(N-n_1-n_2+1)! N!}\right]^{1/2}\, .
\ee
In the large $N$ limit, we can use
\be
\lim_{N\rightarrow \infty} F(n_1,n_2) = \left( \frac{1}{N} \right)^{\frac{1}{2}}\, ,
\ee
and the structure constants (\ref{nbos1},\,\ref{nbos3},\,\ref{nfer1},\,\ref{nfer2}) follow.

The case  (\ref{nraro})  is different, because we have $n_3= n_1 + n_2 -3$, which leads to
\be
\gg=\half \sum_{j=1}^3 \left(n_j-1\right)-c+1 = n_3 -c +1.
\label{graro}
\ee
Here we must distinguish between two cases. If $n_1=2, n_2=n+1$, then $n_3=n$ and the only possible number of colors is $c=n+1=n_3+1$, so $\gg=0$.
The finite $N$ form of (\ref{nraro}) is in this case~\cite{Jevicki:1998bm}
\be
\label{oraro}
\langle O_{n}^{(2,2) \dagger} O_{2}^{(0,0)} O_{n+1}^{(0,0)} \rangle
&=& \left( \frac{N-n}{N(N-1)} \right)^{\frac{1}{2}} \frac{1}{(2n (n+1) )^{1/2}}.
\ee
When both $n_1> 2$ and $n_2>2$, eq.(\ref{graro})  allows $c=n_3,\, \gg=1$ or $c=n_3+1, \, \gg=0$. Here there are contributions from covering surfaces with torus topology
which were not computed so far, and thus the finite $N$ form of the correlator (\ref{nraro}) is not known.

\section{Four-point functions and OPEs}\label{mapsec}

In this section we will compute planar contributions to some four-point functions involving chiral and antichiral operators
and we will explore their OPE limits.

\subsection{The general form of the four-point functions}
As we mentioned above, we are interested in mapping
a covering sphere, $S^2_{cover}$, with coordinate $t$, to the physical sphere $z$, called $S^2_{base}$,
such that the $c$   fields $X_I(z)$, $I = 1, \dots c$ at a generic location $z \in S^2_{base}$   are traded
for a single field $X(t_I(z))$, where $t_I(z) \in \ S^2_{cover} $ are the $c$ pre-images
of the point  $z$. As $z$ approaches a  point $z_i$ where a twist field with  $n_i$ colors is inserted,
we require that $n_i$ of the $c$ preimages of $z$ converge to the same point~$t_i$ on~$S^2_{cover}$.
This implies
\be
t-t_i \sim (z-z_i)^{\frac{1}{n_i}},
\ee
and guarantees that the field $X(t)$ in $S^2_{cover}$ returns to its original position
only after we make $n_i$ full $2\pi$ rotations around point
 $z_i$  on the base sphere $S^2_{base}$.

For a four-point function, we can use the $SL(2,\mathbb{C})$ invariance of $S_{base}$ to fix the twist fields at
\be \label{zi}
z_1 = 0 \, ,\quad z_2 = u \, , \quad z_3 = 1 \, , \quad z_4 = \infty \,,
\ee
and the $SL(2,\mathbb{C})$ invariance of $S_{cover}$ to fix the points where their preimages converge at
\be \label{ti}
t_1 = 0 \, ,\quad t_2 = x \, , \quad t_3 = 1 \, , \quad t_4 = \infty \,.
\ee
Thus the map from $S^2_{cover}$ to $S^2_{base}$ is given by a $c$-sheeted map such that
\be \label{b0}
\lim_{t \rightarrow 0} z(t) &\sim &  b_1 t^{n_1},
\\ \label{b1}
\lim_{t \rightarrow x} z(t) &\sim& u + b_2(t-x)^{n_2},
\\ \label{bx}
\lim_{t \rightarrow 1} z(t) &\sim& 1 + b_3(t-1)^{n_3},
\\
\lim_{t \rightarrow \infty } z(t) &\sim& b_4 t^{n_4} \,.
\label{branch-inf}
\ee
Now, once we obtained a function $z(t)$ satisfying the conditions (\ref{b0})-(\ref{branch-inf})\footnote{
In general there can be several functions satisfying these conditions.
However, in the examples we will consider all the covering maps are obtained from a single $z(t)$,
so this multiplicity will not play any role for us. For a discussion of these issues see \cite{PRR1}.
},  we have not yet fixed the covering map.
 The reason is that~$x$ is a preimage
of $z=u$ and thus has to be chosen to satisfy
\be
u=z(t=x) \,.
\label{hn}
\ee
This equation has in general several, say $M$,  solutions $x_j$ (which in general is different from the number $c$ of preimages of $z(t)$ at
a generic $t \neq x$, because $x$ also appears as a parameter in $z(t)$ for $t \neq x$ ).
The number $M$ is the number of different maps $z(t)$ with the required behavior~(\ref{b0})-(\ref{branch-inf}). 
The problem of counting ramified coverings is referred to as the 
 {\it Hurwitz problem }  in the mathematical literature (see e.g. \cite{Lando:2003gx} for a review).
We will denote each of the covering maps by $z_j(t)$ $(j=1\ldots M)$, and the $c$ inverse maps for each $j$ as $t_{j,I}(z)$ $(I=1 \dots c)$,
and we will have
\be
z_j(x_j) = u   \qquad \qquad j=1\ldots M.
\ee
The crucial observation now, is that  $M$ is precisely the number of terms $H_c$ in (\ref{rc}) corresponding to $\gg=0$.
In other words, $M$ is the number of solutions of the group theory condition (\ref{prodone}) for $p=4$, up to color relabeling, with the constraint
\be
c= \frac12 (n_1+n_2+n_3+n_4) -1 \,.
\ee
For details of this correspondence see again ~\cite{PRR1}.
Therefore, the correlator of gauge invariant operators will be a sum over the $M$ solutions of eq. (\ref{hn}).

Let us define the operators
\be
o_n^{\alpha} &=&  e^{ i \alpha \sum_{I=1}^{n} (\phi_I^1 +  \phi_I^2  + \bar{\phi}_I^1 + \bar{\phi}_I^2 )} {\s}_{n}\, ,
\label{odressedop}
\\
O_n^{\a} &=&  \frac{1}{\sqrt{n\,N!(N-n)!}} \sum_{h\in S(N)}   o_{h^{-1} (12\ldots n)h}^{\a}\,.
\label{dressedop}
\ee
Here $O_n^{\alpha}$ is the normalized and gauge invariant version of $o_n^{\alpha}$.
We will later specialize to $\alpha = \frac{n-1}{2n}$ or $\alpha=\frac{n+1}{2n}$ to get operators of type $O_n^{(0,0)}$ or $O_n^{(2,2)}$.
One can easily generalize the discussion to include operators of type $O_n^{(a,\bar{a})}$.
Consider the four-point function
\be
&& \langle O_{n_4}^{\alpha_4}(z_4, \bar{z}_4) O_{n_3}^{\alpha_3}(z_3,\bar{z}_3) O_{n_2}^{\alpha_2}(z_2,\bar{z}_2) O_{n_1}^{\alpha_1}(z_1,\bar{z}_1) \rangle
_{\gg=0} = G(u, \bar{u})
\label{general4p}
\\
\nn
&& \qquad \qquad \times z_{24}^{-2\Delta_2} \, z_{14}^{\Delta_2 + \Delta_3 -\Delta_1 - \Delta_4 } \, z_{34}^{\Delta_1 + \Delta_2 -\Delta_3 - \Delta_4 }
\, z_{13}^{-\Delta_1 - \Delta_2 -\Delta_3 + \Delta_4 }\times c.c.\, ,
\ee
where
\be
u = \frac{z_{12}z_{34}}{z_{13}z_{24}} \,.
\ee
Expanding the sum over the $S_N$ group in each operator $O_{n_i}^{\alpha_i}$ in (\ref{general4p}), we will have a sum as (\ref{rc}):
\be
&& \langle O_{n_4}^{\alpha_4}(\infty) O_{n_3}^{\alpha_3}(1) O_{n_2}^{\alpha_2}(u) O_{n_1}^{\alpha_1}(0) \rangle_{\gg=0} = G(u, \bar{u})
\nn
\\
&& \qquad \sim \sum_{j=1}^{M} \langle o_{n_4}^{\alpha_4}(\infty) o_{n_3}^{\alpha_3}(1) o_{n_2}^{\alpha_2}(u) o_{n_1}^{\alpha_1}(0) \rangle_{j} =
\sum_{j=1}^{M} G_{j}(u, \bar{u}) \,,
\label{sum4pf}
\ee
where in order to evaluate each term we use a different map $z_j(t)$ to go to the covering surface.
For this we will use the stress-tensor method of Dixon et al. \cite{Dixon:1986qv}.
We compute first the auxiliary function
\be
g_j(z,u) =  \frac{\langle T(z) o_{n_4}^{\alpha_4}(\infty) o_{n_3}^{\alpha_3}(1) o_{n_2}^{\alpha_2}(u,\bar{u}) o_{n_1}^{\alpha_1}(0) \rangle_j}
{\langle o_{n_4}^{\alpha_4}(\infty) o_{n_3}^{\alpha_3}(1) o_{n_2}^{\alpha_2}(u) o_{n_1}^{\alpha_1}(0) \rangle_j} \,\,.
\label{gzu}
\ee
Using now the OPE
\be
T(z)o_{n_2}^{\alpha_2}(u) =\frac{\Delta_2}{(z-u)^2}o_{n_2}^{\alpha_2}(u) +\frac{1}{z-u}\d o_{n_2}^{\alpha_2}(u) +\dots
\ee
we deduce
\be
\d_u \ln G_j(u) =\left\{g_j(z,u)\right\}_{\frac{1}{z-u}} \,,
\label{eqfg}
\ee
where $G_j(u)$ is the contribution to the holomorphic part of $G_j(u, \bar{u})$  and
 on the right hand side we take the coefficient of
$\frac{1}{z-u}$ in the expansion of $g_j(z,u)$. There is a similar anti-holomorphic expression, so that
$G_j(u, \bar{u})=G_j(u)\bar{G}_j(\bar{u})$ up to an overall constant.
The dependence on the index $j$ comes from the fact that
the computation of (\ref{gzu}) is done by mapping, with~$z_j(t)$, all the fields to the covering sphere, where the twist fields disappear.
We need finally to sum over $j$ as in  (\ref{sum4pf}). The relative coefficient is fixed so that the sum is single valued as a function of $u$.
The overall factor is fixed by considering OPE limits.

Before proceeding, note that  the dressing factors
in~(\ref{odressedop}) satisfy
\be
e^{\alpha (\phi_{1} + \ldots  + \phi_n)} \sigma_{(12\ldots n)}
&=& \left(\frac{dt}{dz}\right)^{\Delta} e^{ n \alpha \phi }\, ,
\label{ean}
\ee
where the l.h.s lives in the base sphere and the r.h.s lives in the covering surface.

To obtain the function (\ref{gzu}) we can use that
\be
T(z) &=& -\frac12  \sum_{i=1}^4\d X^i_I(z) \d X^i_I(z)  -\frac12 \sum_{i=1}^2 \d \phi^i_I(z) \d \phi^i_I(z)
\\
&=& -\frac12 \lim_{w \rightarrow z}   \left[  \sum_{i=1}^4 \d X^i_I(z) \d X^i_I(w)
+  \sum_{i=1}^2 \d \phi^i_I(z) \d \phi^i_I(w)    + \frac{6N}{(z-w)^2} \right]\, .
\label{ssing}
\ee
Inserting this expression in~(\ref{gzu}), we must consider the two types of terms.
The terms with~$X^i_I$ in (\ref{ssing}) give
\be
\frac{\langle \d X^i_I(z) \d X^i_I(w) o_{n_4}^{\alpha_4}(\infty) o_{n_3}^{\alpha_3}(1) o_{n_2}^{\alpha_2}(u) o_{n_1}^{\alpha_1}(0) \rangle_j}
{\langle o_{n_4}^{\alpha_4}(\infty) o_{n_3}^{\alpha_3}(1) o_{n_2}^{\alpha_2}(u) o_{n_1}^{\alpha_1}(0) \rangle_j}
= -\frac{t'_{j,I}(z) t'_{j,I}(w)}{(t_{j,I}(z)-t_{j,I}(w))^2} \,,
\ee
and the  terms with $\phi^i_I$  in (\ref{ssing}) give
\be
&& \frac{\langle \d \phi_I^i(z) \d \phi_I^i(w) \, o_{n_4}^{\alpha_4}(\infty) o_{n_3}^{\alpha_3}(1) o_{n_2}^{\alpha_2}(u) o_{n_1}^{\alpha_1}(0) \rangle_j}
{\langle o_{n_4}^{\alpha_4}(\infty) o_{n_3}^{\alpha_3}(1) o_{n_2}^{\alpha_2}(u) o_{n_1}^{\alpha_1}(0) \rangle_j}
\\
\nn
&& \qquad \qquad =
-\frac{t'_{j,I}(z) t'_{j,I}(w)}{(t_{j,I}(z)-t_{j,I}(w))^2}
- (t'_{j,I}(z))^2 \left(  \frac{n_3 \alpha_3}{t_{j,I}(z)-1} +  \frac{n_2 \alpha_2 }{t_{j,I}(z)-x} + \frac{n_1 \alpha_1}{t_{j,I}(z)} \right)^2\, ,
\label{gab}
\ee
where in the second term we took the limit $w \rightarrow  z$ since there is no singularity.

In the sum over $I$ in  (\ref{ssing}), we  only need for $g_j(z,u)$ in (\ref{gzu}) those $n_2$ terms whose index $I$ is
one of the colors appearing in the operator $o_{n_2}^{\alpha_2}$, since only those terms will contribute to the singularity
as~$z \rightarrow u$.
Collecting all the terms, taking the $w \rightarrow z$ limit and subtracting the normal-order singularity as in~(\ref{ssing}), we get finally
\be
g_j(z,u) &=& \frac{6}{12} \sum_{I=1}^{n_2} \{t_{j,I},z \}  + 2  \sum_{I=1}^{n_2}\frac{(t'_{j,I}(z))^2}{2} \left(  \frac{n_3 \alpha_3}{t_{j,I}(z)-1} +  \frac{n_2 \alpha_2 }{t_{j,I}(z)-x} + \frac{n_1 \alpha_1}{t_{j,I}(z)} \right)^2 \,.
\label{gzuf}
\ee
Here
$\{t,z \}$
is the Schwartzian derivative,
\be
\{t, z\} = \frac{t'''}{t'} -\frac32\left(\frac{t''}{t'}  \right)^2 = \left( \frac{t''}{t'} \right)' -\frac12 \left(\frac{t''}{t'} \right)^2\, ,
\ee
and the factor of $2$ in front of the second term comes from the two values $i=1,2$ in $\phi^i$.

Now,  the $n_2$ inverse maps in the sum of (\ref{gzuf})  behave as
\be
t_{j,I}-x_j \sim e^{\frac{ 2 \pi I i}{n_2}}(z-u)^{\frac{1}{n_2}} \qquad \qquad I=1,\ldots,n_2\, .
\label{mapsing}
\ee
Since the terms in (\ref{gzuf}) involve derivatives of $t_{j,I}(z)$, we see that all the $n_2$ terms  contribute to the singularities of $g_j(z,u)$ as $z \rightarrow u$.
Each of these $n_2$ terms has an expansion in powers of $(z-u)^{\frac{1}{n_2}}$, but since $g_j(z,u)$ has no monodromies as $z$ goes around $u$, all the terms in~(\ref{gzuf}) with fractional powers of $(z-u)$ cancel out. Thus we can just take
\be
g_j(z,u) =  \frac{n_2}{2} \{t_{j,I},z \}  + n_2 (t'_{j,I}(z))^2 \left(  \frac{n_3 \alpha_3}{t_{j,I}(z)-1} +  \frac{n_2 \alpha_2 }{t_{j,I}(z)-x} + \frac{n_1 \alpha_1}{t_{j,I}(z)} \right)^2\, ,
\ee
where $t_{j,I}$  is any of the (\ref{mapsing}) maps, and we keep only the terms with integer powers of $(z-u)$ in the expansion of $g_j(z,u)$.

The residue in the $(z-u)^{-1}$ pole in (\ref{eqfg}) will be a function of $x_j$, so it is convenient
to express the l.h.s. of (\ref{eqfg}) as a function of $x_j$ as well, using
\be
\d_u \log G_j(u)=u'(x_j)^{-1}\d_{x_j}\log G_j(u).
\ee
The generic structure of the differential equation for $G_j(z)$ has then the form
\be
u'(x)^{-1}\d_x\log G=\left\{\texttt{A} \left[\left(\frac{t''}{t'}\right)'-\half\left(\frac{t''}{t'}\right)^2\right]
+(t')^2\left[\frac{\texttt{B}}{t-x}+\frac{\texttt{C}}{t}+\frac{\texttt{D}}{t-1}\right]^2\right\}_{\frac{1}{z-u}},
\label{gabcd}
\ee
where
\be
\texttt{A} &=& \frac{n_2}{2},
\\
\texttt{B} &=& n_2^{3/2}\alpha_2 ,
\\
\texttt{C} &=& \sqrt{n_2} n_1 \alpha_1,
\\
\texttt{D} &=& \sqrt{n_2} n_3\alpha_3 ,
\ee
and $t=t_{j,I}(z), x=x_j$.

\subsection{Polynomial maps: a simple example}

To complete the computation we need to construct the explicit map $z(t)$ as a function of the integers
$n_i$. For any $p$-point function, we expect this map to be a quotient of polynomials.
For four-point functions, these polynomials are solutions of Heun's differential equation~\cite{PRR1}.
When the polynomial in the denominator of $z(t)$ has degree zero,  $z(t)$ itself is a polynomial. If we choose the branching of order $n_p$ at $z = \infty$ to correspond to $t = \infty$, this occurs, in a $p$-point function, whenever the branching numbers $n_i$
satisfy
\be
n_p = \sum_{i=1}^{p-1}n_i -p+2\,.
\label{excon}
\ee
This has been shown in detail in~\cite{PRR1} for $p=4$, and the generalization to arbitrary $p$ is immediate.
A correlator whose quantum numbers satisfy the above relation has  {\it only} connected contributions~(cf.(\ref{discoloco})).
Moreover, in this case one can verify that $c=n_p$ and therefore the Riemman-Hurwitz formula (\ref{rh1}) implies that the only covering surface to contribute to the correlator will be a sphere. Note that  the four structure constants (\ref{bos1})-(\ref{fer2}) of the chiral ring satisfy the polynomial condition (\ref{excon}), but
the fifth structure constant (\ref{nraro}) does not.

In this section we will consider a simple polynomial map for a four-point function with branching numbers
\be
n_1&=& n
\\
n_2&=&n_3 =2
\\
n_4&=& n+2\,.
\label{nnn}
\ee
In this case we have $c=n+2$ colors and as expected $\gg=0$.
In Appendix A we present the details of the $z(t)$ function as well as its inverse.
The equation $u=z(x)$ is
\be
u \equiv v(x)=x^{1 + n} \frac{2 + n - n x}{(2+n )x-n}\, ,
\label{umap}
\ee
and has $M=n+2$ solutions. Let us choose a root $x_j$ of (\ref{umap}).
The equation (\ref{gabcd}) can be integrated and the result has the form
\be
\label{gform}
\log G_j(u) &=& \texttt{A} f_A(x_j) + \texttt{B}^2 f_{B^2}(x_j) +
\texttt{C}^2 f_{C^2}(x_j) + \texttt{D}^2 f_{D^2}(x_j)
\\ \nn
&& \qquad + \texttt{2BC} f_{2BC}(x_j)
+ \texttt{2BD} f_{2BD}(x_j)
+ \texttt{2CD} f_{2CD}(x_j)\, ,
\ee
where
\be\label{contribs4p}
f_A &=& \frac{1}{8} \left(-2 \log(-1 + x) - \frac{-2 + n + n^2}{n} \log x +
   2 \frac{-2 + 2 n + n^2}{n (2 + n)} \log(-n + (2+n) x )\right),
\nn
\\	
 f_{B^2} \;&=&\;-\frac{1}{4} \left(\log( x-1) + ( n-1) \log x - \frac{
   n \log( (2  + n) x-n)}{2 + n}\right),
   \nn
 \\
 f_{C^2}\;&=&\;-\frac{(2 + n) \log x - 2 \log( (2+n) x-n)}{2 n (2 + n)},
 \nn
 \\
 f_{2BC}\;&=&\;\frac{ (2 + n) \log x - \log((2+n) x-n)}{2 (2 + n)},\;
f_{D^2}\;=\;-\frac{1}{4} \log( x-1) + \frac{n \log((2+n) x-n)}{8 + 4n},
 \nn
 \\
 f_{2CD}\;&=&\;-\frac{\log4(n -  (2 + n) x)}{2 (2 + n)},\;
 f_{2BD}\;=\;\frac{ (2 + n) \log (x-1) - \log((2+n) x-n)}{2 (2 + n)},\nonumber\\
\ee
with $x=x_j$. Using the expressions in (\ref{gform}), we get $G_j(u)$ and similarly  $\bar{G}_j(\bar{u})$.
We can finally sum over $j$ to get
\be
G(u,\bar{u}) &=& C_4 \sum_{j=1}^{n+2} G_j(u) \bar{G}_j(\bar{u})
\\
&=& C_4 \sum_{j=1}^{n+2}  \left|x_j(u)-1\right|^{2\a}\, \left|x_j(u)\right|^{2\b}\,\left|x_j(u)-\frac{n}{2+n}\right|^{2\gamma}\, ,
\label{fpabg}
\ee
where $C_4$ is a constant and the powers $\a, \gamma$  and $\b$ are,
\be
\a&=&-\frac{1}{4}\left[\texttt{A}+\texttt{B}^2-4\texttt{B}\texttt{D}+\texttt{D}^2\right]
\nn
\\
&=&
-\half\left[\half+4\left(\a_3-\a_2\right)^2-8\a_3\a_2\right] \,,
\label{alpha}
\\
\b&=&-\frac{1}{4}\left[\frac{-2+n+n^2}{2n}\texttt{A}+(n-1)\texttt{B}^2-4\texttt{BC}
+\frac{2}{n}\texttt{C}^2\right]
\nn
\\&=&-\left[\frac{-2+n+n^2}{8n}+n\left(2\a_2^2-4\a_2\a_1+\a_1^2\right)-2\a_2^2\right] \,,
\label{beta}
\\
\gamma&=& \frac{1}{4}\left[\frac{-2 + 2n  + n^2}{n (2 + n)}\texttt{A}+\frac{n}{n+2}\texttt{B}^2-\frac{4}{n+2}\texttt{B}\left(\texttt{C}+
\texttt{D}\right)
%+\frac{4}{n(n+2)}\texttt{C}^2+\frac{n}{n+2}\texttt{D}^2-\frac{4}{n+2}\texttt{CD
+\frac{4}{n+2}\left(\frac{\texttt{C}}{\sqrt{n}}-\frac{\sqrt{n}}{2}\texttt{D}\right)^2\right]
\nn
\\
&=&\frac{1}{4n(n+2)}\left[n^2+2n-2+8 n^2 \left(\a_1^2+\a_2^2+\a_3^2\right)-16\left(\a_1 n^2 \left(\a_2+\a_3\right	)+2 n \a_2\a_3\right)\right]\,.
\nn
\\
\label{gamma}
\ee

\subsection{Some non-extremal four-point functions and their OPEs}
We have now all the ingredients to compute some examples explicitly, using the polynomial map we introduced above.
Consider
\be
\langle O_{n+2}^{(0,0) \dagger}(\infty) O_{2}^{(0,0) }(1)  O_{2}^{(0,0) \dagger}(u,\bar{u})  O_{n}^{(2,2)}(0)  \rangle = G(u,\bar{u})\,.
\label{ffpf}
\ee
The quantum numbers are $\alpha_1=\frac{n+1}{2n}, \alpha_2= - \alpha_3= - 1/4$, from which we get
%$ \texttt{B} = -\texttt{D} = - \frac{1}{\sqrt{2}}, \texttt{C} =   \frac{(n+1)}{\sqrt{2}}$, and~$\texttt{A}=1$.
%This in turn implies
  $ \alpha = -1, \beta = -n-1,  \gamma = 1$, and thus, from (\ref{fpabg}),
\be
G(u,\bar{u})  = C_4  \sum_{i=1}^{n+2} \left|x_i(u)-1\right|^{-2} \, \left|x_i(u)\right|^{-2n-2} \,\left|x_i(u)-\frac{n}{2+n}\right|^{2}\,.
\label{gzfp}
\ee
It only remains to find $C_4$, which can be done by considering OPE limits.
Note that since the polynomial map leads to sphere contributions only, we expect to find the finite $N$ expression for $C_4$.

Consider the limit $u\to 0$.
The equation $u(x)=0$ has $n+1$ roots with $x=0$ and one root with $x=\frac{2+n}{n}$.
Taking the former case, we obtain
\be\label{ubeh1}
u=-\frac{2+n}{n}x^{n+1} + O(x^{n+2}).
\ee
Inserting this into (\ref{gzfp}) we get that
\be
G(u,\bar u)|_{u\to 0} = C_4 (n+1) |u|^{-2} + O(u^{-1})+ O(\bar{u}^{-1}) + O(1)\, ,
\label{uex}
\ee
where the factor $(n+1)$ comes from the number of terms in (\ref{gzfp}) which have the behavior~\eqref{ubeh1}.
The terms of order $O(1)$ include the leading contribution to (\ref{gzfp}) from the root~$x=\frac{2+n}{n}$, which is a constant.

Now, in the limit $u\to 0$ we have
\be
O_{2}^{(0,0) \dagger}(u, \bar{u})  O_{n}^{(2,2)}(0) \sim  \frac{ C_1 \,O_{n-1}^{(2,2)} (0)}{|u|^2} + \frac{C_2 O_{n+1}^{(0,0)}(0)}{|u|^2}\, .
\ee
The two operators appearing in this OPE happen to be chiral because (\ref{uex}) fixes their conformal dimension to be $\Delta= \Delta_{2}^{0} + \Delta_{n}^2-1= \frac{n}{2}$, and their charge is  $Q=\frac{n}{2}$ from charge conservation, but in general the OPE of a chiral operator with an antichiral operator is neither chiral nor antichiral.
Inserting this OPE in (\ref{ffpf}), we see that only the second term can survive due to the length of the cycles. Therefore the OPE limit yields
\be
C_4 (n+1) = \langle O_{n+2}^{(0,0) \dagger} O_{2}^{(0,0)} O_{n+1}^{(0,0)} \rangle \langle O_{n}^{(2,2)} O_{2}^{(0,0) \dagger}O_{n+1}^{(0,0) \dagger} \rangle
\, ,
\ee
and using the  structure constants  (\ref{bos1}) and (\ref{oraro}) (the (a,a) ring has the same structure constants as the (c,c) ring), we get the finite $N$ result
\be
C_4 = \frac{(n+2)^{3/2}}{2 (n+1)^2  n^{1/2}}
\sqrt{\frac{(N-n)(N-n-1)}{N^2(N-1)^2}} \,.
\label{c4}
\ee

\vskip .3cm
\ni
We can verify this expression  for $C_4$ by reobtaining it in another limit. When  $u \rightarrow \infty$, we get
$n+1$ solutions with $x\to\infty$ and a single solution with $x\sim \frac{n}{2+n}$. In the former case one explicitly obtains that
\be
u \sim - \frac{n}{n+2} x^{n+1} \,.
\ee
Inserting this into (\ref{gzfp}) gives
\be
G(u,\bar u)|_{u\to \infty} = C_4 (n+1) \frac{n^2}{(n+2)^2} |u|^{-2} +O(1)\, ,
\label{olpc}
\ee
where the $(n+1)$ is again the number of terms in (\ref{gzfp}) which diverge in the OPE limit.
The OPE leads again to a chiral intermediate state $O_{n+1}^{(2,2)}$ and thus we get
\be
C_4 \frac{n^2(n+1)}{(n+2)^2} =  \langle  O_{2}^{(0,0) \dagger} O_{n+2}^{(0,0) \dagger}  O_{n+1}^{(2,2)} \rangle
\langle O_{n+1}^{(2,2) \dagger} O_{2}^{(0,0)} O_{n}^{(2,2)} \rangle\, .
\ee
Inserting the structure constants (\ref{oraro}) and (\ref{bos3}), we get for $C_4$ exactly the same expression as in (\ref{c4}).
One could try to further take the limit $u\rightarrow 1$, but in this case the intermediate state is neither chiral nor anti-chiral, so we cannot use the structure constants of the chiral ring.

\vskip .7cm

Consider next the four-point function
\be
\langle O_{n+2}^{(0,0) \dagger}(\infty) O_{2}^{(0,0) \dagger }(1)  O_{2}^{(0,0) }(u)  O_{n}^{(2,2)}(0)  \rangle\, .
\label{ffps}
\ee
This correlator is obtained from (\ref{ffpf}) by interchanging the positions of $z_2=u$ with $z_3=1$.
Using (\ref{general4p}), this implies that (\ref{ffps}) is equal to
\be
\langle O_{n+2}^{(0,0) \dagger}(\infty) O_{2}^{(0,0) }(1)  O_{2}^{(0,0) \dagger}(1/u,1/\bar{u})  O_{n}^{(2,2)}(0)  \rangle  \times |u|^{-2} \,.
\ee
In general, $SL(2,\mathbb{C})$ transformations in the $u$ sphere are not simple in the $x$ sphere, but for our particular map (\ref{umap}), we have
\be
1/v(x) = v(1/x)\,,
\ee
and therefore
\be
\langle O_{n+2}^{(0,0) \dagger}(\infty) O_{2}^{(0,0) \dagger }(1)  O_{2}^{(0,0) }(u)  O_{n}^{(2,2)}(0)  \rangle  =
C_4  \sum_{j=1}^{n+2} \left|x_i(u)-1\right|^{-2} \,  \, \left|x_i(u)-\frac{n}{2+n}\right|^{2}\, .
\ee
This result can be verified by computing the coefficients $\alpha, \beta, \gamma$ explicitly.
Note that there is no singularity as $u \rightarrow 0$, since this corresponds to the OPE of two chiral operators.

\section{Extremal correlators  and Hurwitz numbers}\label{extremal}

Armed with this experience, we are ready to  tackle the computation of extremal correlators.

\subsection{Extremal four-point functions and double-cycle operators}
Consider the following extremal four-point functions
\be
\langle O_{n+2}^{(0,0) \dagger}(\infty) O_{2}^{(0,0) }(1)  O_{2}^{(0,0) }(u,\bar{u})  O_{n}^{(0,0)}(0)  \rangle = G_1(u,\bar{u})\,,
\\
\langle O_{n+2}^{(2,2) \dagger}(\infty) O_{2}^{(2,2) }(1)  O_{2}^{(0,0) }(u,\bar{u})  O_{n}^{(0,0)}(0)  \rangle = G_2(u,\bar{u})\,,
\label{extrem2}
\ee
which we can compute with the same map of the previous section.
The quantum numbers for $G_1$ are $\a_1 = \frac{n-1}{2n}, \a_2=\a_3=\frac14$. 
Using these values in (\ref{alpha})-(\ref{gamma}) gives $\a=\b=\gamma=0$.
Similarly, for $G_2$ we have $\a_1 = \frac{n-1}{2n}, \a_2 =\frac14, \a_3=\frac34$ and we get again $\a=\b=\gamma=0$.
From~(\ref{fpabg}), this means that both $G_1$ and $G_2$ are constants.
We get thus a  nice check on  our formulas for~$\a,\b,\gamma$, since this is precisely what we expect:
the limits~$u \rightarrow 0,1$ correspond to OPEs of chiral operators, so  $G_1(u)$ and $G_2(u)$
should  have no singularities at~$u= 0,1$. But these are the only singularities we expect in~$G_1(u),G_2(u)$.
So  $G_1(u),G_2(u)$ are  meromorphic functions with no singularities and therefore should be constants.
This result is of course general, so we will have
\be
\langle O_{n_4}^{(0,0) \dagger}(\infty) O_{n_3}^{(0,0) }(1)  O_{n_2}^{(0,0) }(u,\bar{u})  O_{n_1}^{(0,0)}(0)  \rangle = C_4\, ,
\label{extremlim}
\ee
with $C_4$ a constant and
\be
n_4=n_3 + n_2 + n_1 -2\, ,
\label{n4sum}
\ee
from charge conservation, which is the familiar polynomial condition~(\ref{excon}).

It only remains to determine $C_4$ from OPE limits. But here things are subtler than for non-extremal correlators, because the
OPE of two chiral operators has no singularities. Taking for example $O_{n_{2}}^{(0,0)}$   and  $O_{n_{1}}^{(0,0)}$, we get
\be
O_{n_{2}}^{(0,0)} O_{n_{1}}^{(0,0)} &=&  C_3 O_{\tilde{n}}^{(0,0)} + C'_3 O_{\tilde{n}-2}^{(2,2)}
\label{extremalope}
\\
\nn
&& + \sum_{i=2}^{\tilde{n}-1} \left( D_3(i)  O_{(i,\tilde{n} -i+1)}^{(0,0),(0,0)}
+ D_3'(i) O_{(i-1,\tilde{n} -i)}^{(2,2),(0,0)} 
+ \cdots \right)
+ \cdots
\ee
where
\be
\tilde{n} = n_1 +n_2 -1 \, .
\label{tn}
\ee The operators in the parentheses in the second line of (\ref{extremalope}) are double cycle operators.
They are always present in the OPE of single cycle operators and correspond, for instance, to a product of two single cycles without common colors.
Since such a product has no singularities, we ignored these operators in the previous section.
But in the OPE of two chiral operators they appear on par with the single cycle
operators.

The problem now is that we do not know the structure constants $D_3, D_3'$ (and other structure constants including multi-cycle
operators) in~(\ref{extremalope}),
so we cannot use this OPE to determine $C_4$ in~(\ref{extremlim}). Of course, in principle these structure constants can be computed using, e.g., the techniques of \cite{Lunin:2001pw}. But this has not been done so far, and as we will see below, it is not necessary.

The presence of the double cycle terms cannot be avoided even in the large $N$ limit.
To see this, insert the OPE (\ref{extremalope}) into (\ref{extremlim}). This gives
\be
C_4 &=& C_3 \langle O_{n_4}^{(0,0) \dagger} O_{n_3}^{(0,0)} O_{\tilde{n}}^{(0,0)} \rangle
+ \sum_{i=2}^{\tilde{n}-1}   D_3(i)  \langle O_{n_4}^{(0,0) \dagger} O_{n_3}^{(0,0)}  O_{(i,\tilde{n} -i+1)}^{(0,0),(0,0)} \rangle
\\
\label{c4e}
&=& \langle O_{\tilde{n}}^{(0,0) \dagger} O_{n_{2}}^{(0,0)} O_{n_{1}}^{(0,0)} \rangle  \langle O_{n_4}^{(0,0) \dagger} O_{n_3}^{(0,0)} O_{\tilde{n}}^{(0,0)} \rangle
\\
\nn
&& \qquad + \sum_{i=2}^{\tilde{n}-1}   \langle O_{(i,\tilde{n} -i+1)}^{(0,0),(0,0) \dagger} O_{n_{2}}^{(0,0)} O_{n_{1}}^{(0,0)} \rangle
\langle O_{n_4}^{(0,0) \dagger} O_{n_3}^{(0,0)}  O_{(i,\tilde{n} -i+1)}^{(0,0),(0,0)} \rangle \, .
\ee
Remember from (\ref{largen}) that the three-point functions of single cycle operators scale as~$N^{-\frac12}$, so the first term in (\ref{c4e}) scales as $N^{-1}$.
Using the same combinatorial arguments used to obtain (\ref{largen}) (see e.g. \cite{PRR1}), it is easy to show that at large $N$
\be
\langle O_{(n_2,n_1) }^{(0,0),(0,0) \dagger} O_{n_{2}}^{(0,0)} O_{n_{1}}^{(0,0)} \rangle & \sim &   1\, ,
\label{discon}
\\
\langle O_{n_4}^{(0,0) \dagger} O_{n_3}^{(0,0)}  O_{(n_2,n_1)}^{(0,0),(0,0)} \rangle & \sim & \frac{1}{N} \,.
\ee
In particular, the correlator (\ref{discon}) only receives contributions from its  disconnected terms (see (\ref{discoloco})).
It follows then that in the sum in (\ref{c4e}) there is always a term that scales as~$N^{-1}$, as the first term, and thefore cannot be discarded.

A similar situation occurs in a free gauge theory~\cite{D'Hoker:1999ea}. Consider  the  following correlator in free $U(N)$ gauge theory
\be
I^{(YM)}_4=\langle \Tr{\bar Z}^J(x_0)\Tr Z^{J_1}(x_1)\Tr Z^{J_2}(x_2)\Tr Z^{J_3}(x_3)\rangle,
\ee where $J=J_1+J_2+J_3$. Let us take the $OPE$ limit $x_1\to x_0$. The leading contributions to this are
\be\label{YMOPE}
I^{(YM)}_4=&&
\frac{1}{|x_0-x_1|^{2J_1}}\biggl[\langle \Tr{\bar Z}^{J-J_1}(x_0)\Tr Z^{J_2}(x_2)\Tr Z^{J_3}(x_3)\rangle\langle \Tr Z^{J-J_1}\Tr Z^{J_1}\Tr{\bar Z}^{J}\rangle +\\
&&+\langle \left(\Tr{\bar Z}^{J_2}\Tr{\bar Z}^{J_3}\right)(x_0)\Tr Z^{J_2}(x_2)\Tr Z^{J_3}(x_3)\rangle\langle \left(\Tr Z^{J_2}\Tr Z^{J_3}\right)\Tr Z^{J_1}Tr{\bar Z}^{J}\rangle+\dots\biggr]\nonumber .
\ee In general the second term coming from a double-trace state is subleading in $1/N$, but here the two terms are of the same order. To see this
we  normalize the operators so that the two point functions will be $O(N^0)$ and thus the planar $s$-point functions of single traces behave as
$N^{2-s}$. The first term in \eqref{YMOPE} is of order $N^{-2}$ and the second naively is of order $N^{-4}$. However, the leading contribution
to the  three-point function $$\langle\left(\Tr{\bar Z}^{J_2}\Tr{\bar Z}^{J_3}\right)(x_0)\Tr Z^{J_2}(x_2)\Tr Z^{J_3}(x_3)\rangle$$
comes from the disconnected diagrams which
gives a scaling of $N^{-2}$ to the second term in \eqref{YMOPE}. Thus the two terms have the same order in $1/N$ and both have to be counted.

\vskip .5cm
\subsection{The $\e$-deformation}
To bypass the problem of determining the structure constants involving two-cycle states we would like to deform
slightly our  extremal correlators, so as to  split single cycle from  double cycle terms in the OPE (\ref{extremalope}).
We can achieve this using the following trick.
Let us change infinitesimally the momentum of, say, $O_{n_{2}}^{(0,0)}$ and $O_{n_{3}}^{(0,0)}$, as  $\a_2 \rightarrow \alpha_2 = \frac{n_2-1}{2n_2}-\e$
and $\a_3 \rightarrow \alpha_3 = \frac{n_3-1}{2n_3}+\e$.
The OPE (\ref{extremalope}) is now\footnote{Note that since the radius of the bosons $\phi_I^i$ was fixed from bosonizing the fermions,
 a change of their momentum by $\e$  is actually a change in the radius away from the fermionization point.
This in turn implies that we should deform the momenta of all the operators in the correlator.
But we can ignore this subtlety, since in the sphere the correlators are analytic functions of the external momenta and there is no other 
dependance on the compactification radius.}
\be
O(\e)_{n_{2}}^{(0,0)}(u, \bar{u}) O_{n_{1}}^{(0,0)}(0)  &=&  \frac{C_3(\e) O(\e)_{\tilde{n}}^{(0,0)}(0) } {|u|^{2 \a_1 \e}}
+ \frac{C'_3(\e) O(\e)_{\tilde{n}-2}^{(2,2)} (0)} {|u|^{2 \a_1 \e}}
\label{deformedope}
\\
\nn
&& + \sum_{i=2}^{\tilde{n}-1} \left( D_3(i,\e)  O(\e)_{(i,\tilde{n} -i)}^{(0,0),(0,0)}
+ D_3'(i,\e) O(\e)_{(i-1,\tilde{n} -i)}^{(2,2),(0,0)} % + D_3''(i)O_{(i-1,\tilde{n} -i)}^{(2,2),(2,2)}
+ \cdots \right)
+ \cdots
\ee
The four-point function (\ref{extremlim}) will now have a form similar to (\ref{fpabg})
\be
\langle O_{n_4}^{(0,0) \dagger}(\infty) O(-\e)_{n_3}^{(0,0) }(1)  O(\e)_{n_2}^{(0,0) }(u,\bar{u})  O_{n_1}^{(0,0)}(0) \rangle
=   \frac{C_4}{H_4}\sum_{j=1}^{H_4} \tilde C_j(\e)\prod_{i=1}^{t} |x_{j}(u)-q_t|^{2 \b_t(\e)}\, ,\nonumber\\
\label{deformed4pf}
\ee
such that $\lim_{\e \to 0} \b_t(\e) =0$ and $\lim_{\e \to 0} \tilde C_j(\e) =1$. The sum is over the number of maps from the covering surface, which
 for four-point functions satisfying~(\ref{n4sum}) is $H_4=n_4$ (see e.g. Section 3.3 in \cite{PRR1}).

We can now take the limit $u \to 0$. In~\cite{PRR1} we proved
that the number of terms in the sum over $j$ in (\ref{deformed4pf}) which contribute to the leading singularity is precisely the number $\tilde{n}$ defined in~(\ref{tn}).
Thus for $\tilde{n}$ terms in (\ref{deformed4pf}) we will have
\be
\prod_{i=1}^{t} |x_{j}(u)-q_t|^{2 \b_t(\e)} \to \frac{c_j^{\e}}{|u|^{2 \a_1 \e}}\, ,
\ee
with $c_j$ some constant. Inserting now the OPE (\ref{deformedope}) in (\ref{deformed4pf}), and equating the terms with
leading singularity $|u|^{-2 \a_1 \e}$, we get
\be
\frac{C_4}{n_4}\sum_{j=1}^{\tilde{n}} c_j^{\e} = \langle O_{n_4}^{(0,0) \dagger} O(-\e)_{n_{3}}^{(0,0)}   O(\e)_{\tilde{n}}^{(0,0)} \rangle
\langle O(\e)_{\tilde{n}}^{(0,0) \dagger} O(\e)_{n_{2}}^{(0,0)}   O_{{n}_1}^{(0,0)} \rangle\, .
\ee
We can now safely take the limit $\e \to 0$, and we get,
\be
C_4 \frac{\tilde n}{n_4} = \langle O_{n_4}^{(0,0) \dagger} O_{n_{3}}^{(0,0)}   O_{\tilde{n}}^{(0,0)} \rangle
\langle O_{\tilde{n}}^{(0,0) \dagger} O_{n_{2}}^{(0,0)}   O_{{n}_1}^{(0,0)} \rangle\, .
\label{chiralopefact}
\ee
Note that if we keep only the single-cycle terms in the OPE (\ref{extremalope}), we would
get an expression similar to (\ref{chiralopefact}) but without the factor $\frac{\tilde{n}} {n_4}$. Thus the combined effect of the double cycle
terms in (\ref{extremalope}) is precisely to add this factor.

Inserting now the three-point functions (\ref{bos1}) into (\ref{chiralopefact}), gives finally
\be
C_4 = F_4(n_i) \frac{n_4^{5/2}}{(n_1 n_2 n_3)^{1/2}}\, ,
\label{c4ext}
\ee
where
\be
F_4(n_i) = \left[ \frac{(N-n_1)! (N-n_2)! (N-n_3)! }{(N-n_4)! (N!)^2} \right]^{1/2}\, .
\ee
Note that the final expression (\ref{c4ext}) does not depend on the length $\tilde{n}$ of the intermediate state.
In particular we can repeat the derivation of $C_4$ using any other OPE limit and obtain the same result.

This computation can be easily extended to other extremal correlators.
Using the chiral fusion rules (\ref{fusionrules}),
one can verify that all possible extremal correlators can be obtained by adding operators  of type $(0,0)$ to the structure
constants of the chiral ring.

When adding one operator of type $(0,0)$ to the structure constants (\ref{bos1},\,\ref{bos3},\,\ref{fer1},\,\ref{fer2}) cases, we
get again the polynomial relation (\ref{n4sum}). Thus the counting of total and divergent terms is the same as in the above case,
and we get
\be
\langle O_{n_{4}}^{(0,0) \dagger} O_{n_{3}}^{(0,0)} O_{n_{2}}^{(0,0)}  O_{n_{1}}^{(0,0)} \rangle
&=& F_4(n_i) \frac{n_4^{5/2}}{(n_1 n_2 n_3)^{1/2}}\, ,
\\
\langle O_{n_{4}}^{(2,2) \dagger} O_{n_{3}}^{(2,2)} O_{n_{2}}^{(0,0)}  O_{n_{1}}^{(0,0)} \rangle
&=& F_4(n_i) \frac{n_{3}^{3/2} n_4^{1/2}} {( n_{2} n_1 )^{1/2}}\, ,
\\
\langle O_{n_{4}}^{(b,\bar{b}) \dagger} O_{n_{3}}^{(a,\bar{a})} O_{n_{2}}^{(0,0)}   O_{n_{1}}^{(0,0)} \rangle
&=& \delta^{ab} \delta^{\bar{a}\bar{b}} F_4(n_i) \frac{n_4^{3/2} n_{3}^{1/2} }{(n_{2}  n_1 )^{1/2}}\, ,
\\
\langle O_{n_{4}}^{(2,2) \dagger} O_{n_{3}}^{(a,\bar{a})} O_{n_{2}}^{(b,\bar{b})} O_{n_{1}}^{(0,0)}  \rangle
&=&  \e^{ab} \e^{\bar{a}\bar{b}} F_4(n_i) \frac{(n_{4} n_3 n_{2})^{1/2}}{n_1 ^{1/2}}\, .
\ee

\vskip 1cm

\subsection{Extremal non-polynomial four-point function}
The correlator
\be
\langle O_{n_{4}}^{(2,2) \dagger} O_{n_{3}}^{(0,0)} O_{n_{2}}^{(0,0)}   O_{n_{1}}^{(0,0)} \rangle
\label{exraro}
\ee
is different from the   cases studied above because the conservation of charge here leads to
\be
n_4 = n_1+n_2+n_3-4\, .
\ee
The corresponding map is not polynomial (cf.(\ref{n4sum})), but rather
the quotient of a polynomial and a monomial. Of course one can still build the map
following the prescription developed in~\cite{PRR1},\footnote{For the case $n_1=n_4=n, n_2=n_3=2$, see~\cite{Lunin:2000yv,PRR1}.}
but the $\e$-deformation technique that we introduced does not need the actual map. All we need is the total number of different maps and the
number of terms in the sum over maps which diverge in the possible OPEs. But this information can be obtained from the symmetric-group theoretical version
of the map counting, and then it reduces to a combinatorial problem.

Note that in the previous cases, when we took the OPE (\ref{extremalope}), the operator $O^{(2,2)}_{\tilde{n}-2}$ did not survive inside the correlation function. But in the four-point function (\ref{exraro}) it does and therefore the $\e$-deformed four-point function is, to leading order,
\be
\langle O_{n_{4}}^{(2,2) \dagger} O(-\e)_{n_{3}}^{(0,0)} O(\e)_{n_{2}}^{(0,0)}   O_{n_{1}}^{(0,0)} \rangle
&=& |u|^{-2\e \a_1} \times
\\
\nn
&&
\!\!\!\!\!\!\!\!\!\!\!\!\!\!\!\!\!\!\!\!\!\!\!\!\!\!\!\!\!\!\!\!\!\!\!\!\!\!\!\!\!\!\!\!\!\!\!\!\!\!\!\!\!\!\!\!\!\!\!\!\!
\!\!\!\!\!\!\!\!\!\!\!\!\!\!\!\!
\left[  \langle O_{n_{4}}^{(2,2) \dagger}  O(-\e)_{n_{3}}^{(0,0)} O(\e)_{n_1+n_2-1}^{(0,0)} \rangle  C_3(\e)
+ \,
\langle O_{n_{4}}^{(2,2) \dagger}  O(-\e)_{n_{3}}^{(0,0)} O(\e)_{n_1+n_2-3}^{(2,2)} \rangle
C_3'(\e) \right]
\\
\label{opetwot}
&=&
|u|^{-2\e \a_1} \times
\nn
\\
&&
\!\!\!\!\!\!\!\!\!\!\!\!\!\!\!\!\!\!\!\!\!\!\!\!\!\!\!\!\!\!\!\!\!\!\!\!\!\!\!\!\!\!\!\!\!\!\!\!\!\!\!\!\!\!\!\!\!\!\!\!\!
\!\!\!\!\!\!\!\!\!\!\!\!\!\!\!\!\!\!\!\!\!\!\!\!\!\!\!\!\!\!\!\!
\left[
\langle O_{n_{4}}^{(2,2) \dagger} O(-\e)_{n_{3}}^{(0,0)} O(\e)_{n_1+n_2-1}^{(0,0)} \rangle
\langle O(\e)_{n_1+n_2-1}^{(0,0) \dagger} O_{n_1}^{(0,0)} O(\e)_{n_2}^{(0,0)}\rangle
\right.
\\
\nn
&&
\!\!\!\!\!\!\!\!\!\!\!\!\!\!\!\!\!\!\!\!\!\!\!\!\!\!\!\!\!\!\!\!\!\!\!\!\!\!\!\!\!\!\!\!\!
\left.
+
\langle O_{n_{4}}^{(2,2) \dagger} O(-\e)_{n_{3}}^{(0,0)} O(\e)_{n_1+n_2-3}^{(2,2)}  \rangle
\langle O(\e)_{n_1+n_2-3}^{(2,2) \dagger} O_{n_1}^{(0,0)} O(\e)_{n_2}^{(0,0)}\rangle
\right]\, .
\ee
On the other hand, following the same logic as in the previous section, the above expression is equal to
\be
|u|^{-2\e \a_1}  \frac{C_4}{H_4}\left[ \sum_{j=1}^{\tilde{n}_b} c_j^{\e} + \sum_{j=1}^{\tilde{n}_a} {\tilde c}_j^{\e} \right]\, ,
\label{ulalo}
\ee
where $H_4$ is the total number of mappings. We have separated the  terms contributing to the OPE singularity into two sums.
In the first $\tilde n_b$ terms, the operators $O_{n_2}^{(0,0)}$ and  $O_{n_1}^{(0,0)}$  share one color, and in the other $\tilde n_a$  terms
they share two colors. These two sums clearly correspond to the two terms in the r.h.s. of~(\ref{opetwot}).
In Appendix \ref{nonpol}
we compute these numbers to be
\be
\tilde n_a=n_1+n_2-3,\quad \tilde n_b=n_1+n_2-1,\quad H_4=2\, n_4.
\label{dnonpol}
\ee
We can equate either of the two terms in (\ref{opetwot}) and (\ref{ulalo}). Taking $\e\to 0$ gives the equations
\be
C_4\,\frac{\tilde n_a}{H_4}&=&\langle O_{n_{4}}^{(2,2) \dagger} O_{n_1+n_2-3}^{(2,2)} O_{n_{3}}^{(0,0)}\rangle
\langle O_{n_1+n_2-3}^{(2,2) \dagger} O_{n_1}^{(0,0)} O_{n_2}^{(0,0)}\rangle,\label{c4a}\\
C_4\,\frac{\tilde n_b}{H_4}&=&\langle O_{n_{4}}^{(2,2) \dagger} O_{n_1+n_2-1}^{(0,0)} O_{n_{3}}^{(0,0)}\rangle
\langle O_{n_1+n_2-1}^{(0,0) \dagger} O_{n_1}^{(0,0)} O_{n_2}^{(0,0)}\rangle,\label{c4b}
\ee
and plunging the relevant structure constants we get, at large $N$,
\be\label{extraextC}
C_4=\frac{2}{N}\frac{(n_4)^{1/2}}{\left(n_1n_2n_3\right)^{1/2}}.
\ee
The fact that we get the same result by equating separately the first and second terms in (\ref{opetwot}) and (\ref{ulalo})
is a non trivial check of the procedure.
Note that to extend \eqref{extraextC}  to finite $N$ we would have to compute torus contributions.

\vskip 1cm

\subsection{Extremal polynomial  $p$-point functions}

We can now generalize the above results  to extremal $p$-point
 functions. As in the case of four-point functions, there are four polynomial correlators and one non-polynomial.
We will consider only the former for simplicity.

The coordinate dependence  of extremal $p$-point correlators is
 \be
 \langle O_{n_{p}}^{(0,0) \dagger}(z_p)        O_{n_{p-1}}^{(0,0)}(z_{p-1})  \ldots   O_{n_{1}}^{(0,0)}(z_1) \rangle
 = C_p  \prod_{i=1}^{p-1} |z_i-z_p|^{-4 \Delta_i }\, ,
 \label{allchiralf}
 \ee
 where  $C_p$ is a constant.
  This is fixed by the absence of singularities when the chiral operators approach each other and by requiring invariance under global conformal transformations. We will  put $O_{n_{p}}^{(0,0) \dagger}(z_p)$ at $z_p =\infty$, so this expression becomes just~$C_p$.

To determine  $C_p$ from OPE limits, we  proceed as above. In order to avoid the multi-cycle terms in the OPEs we deform again  the
momentum of $O_{n_{2}}^{(0,0)}$ by  $\e$ and the momentum of $O_{n_{3}}^{(0,0)}$ by $-\e$.
The correlator becomes the usual sum over all the maps from the covering surface
\be
\langle O_{n_{p}}^{(0,0) \dagger}        O_{n_{p-1}}^{(0,0)}  \ldots   O(\e)_{n_{2}}^{(0,0)} O_{n_{1}}^{(0,0)} \rangle =
\frac{C_p}{H_p}\sum_{j=1}^{H_p} |k\left(x_{j}(z_1,z_2,\ldots)\right)|^{2\e}\, .
\label{hpsum}
\ee
Note that charge conservation in (\ref{allchiralf}) implies the polynomial condition
\be
n_p = \sum_{i=1}^{p-1}n_i -p+2 \,.
\label{extremalp}
\ee
Since  $c=n_p$ for polynomial maps, the number $H_p$ of terms in (\ref{hpsum}) is the number of $n_p$-sheeted covering maps from $S^2_{cover}$ to  $S^2_{base}$ with $p$ branching points, with branching numbers $n_1,n_2,\ldots n_p$.
The problem of determining $H_p$ is well known in the mathematical literature on branched coverings, and its solution when  (\ref{extremalp}) holds is  \cite{Lando:2003gx,Lando2}
\be\label{hurext} H_p=n_p^{p-3}.
\ee
For $p=4$ we proved this result in~\cite{PRR1}, and for $p=5$ we present a proof in Appendix~\ref{ntilde} as an illustration of the diagrammatic description of symmetric products we introduced in~\cite{PRR1}.

The details of the functions $k(x_{j}(z_1,z_2,\ldots))$ in~(\ref{hpsum}) have not been worked out for $p>4$, but they are not important for us. We are only interested in the fact that as  $z_2 \to z_1$, a certain number~$\tilde{n}$ of terms in (\ref{hpsum}) will behave as
\be
|k(x_{j}(z_1,z_2,\ldots)|^{2\e}  \to \frac{c_j^{\e}}{|z_{12}|^{2 \e \a_1}}.
\ee
This number $\tilde{n}$ counts how many terms in (\ref{hpsum}) contribute to the singularity of the deformed OPE (\ref{deformedope}).
In Appendix~\ref{opecount} we prove that~$\tilde{n}= (n_1+n_2-1)n_p^{p-4}$.\,
Therefore, in the limit~$z_2 \to z_1$, the coefficients of the leading $|z_{12}|^{-2 \e \a_1}$ singularity at both sides of (\ref{hpsum}) satisfy
\be
\frac{C_p}{n_p^{p-3}}\sum_{j=1}^{(n_1+n_2-1)n_p^{p-4}} c_j^{\e} =
\langle O(\e)_{n_1+n_2-1}^{(0,0) \dagger} O(\e)_{n_{2}}^{(0,0)}   O_{{n}_1}^{(0,0)} \rangle C(\e)_{p-1},
\label{trulala}
\ee
where
\be
C(\e)_{p-1} = \langle O_{n_p}^{(0,0) \dagger} O_{n_{p-1}}^{(0,0)} \ldots   O(-\e)_{n_3}^{(0,0)} O(\e)_{n_1+n_2-1}^{(0,0)} \rangle.
\ee
We can take now the limit $\e \to 0$ in both sides of (\ref{trulala}) to get
\be
C_p \frac{(n_1+n_2-1) }{n_p} &=&
\langle O_{n_1+n_2-1}^{(0,0) \dagger} O_{n_{2}}^{(0,0)}   O_{{n}_1}^{(0,0)} \rangle C_{p-1}
\\
&=& F(n_1,n_2) \frac{(n_1+n_2-1)^{3/2}}{(n_1 n_2)^{1/2}} C_{p-1},
\ee
or equivalently
\be
C_p = F(n_1,n_2)  \frac{ n_p (n_1+n_2-1)^{1/2}}{(n_1 n_2)^{1/2}} C_{p-1}.
\ee
Iterating this recursion relation $p-3$ times, we get finally
\be
\langle O_{n_{p}}^{(0,0) \dagger} O_{n_{p-1}}^{(0,0)}  \ldots   O_{n_{2}}^{(0,0)} O_{n_{1}}^{(0,0)} \rangle = F_p(n_i)  \frac{ (n_p)^{p-\nicefrac32} }{(n_1 n_2\ldots n_{p-1})^{1/2}},
\ee
where
\be
F_p(n_i) = \left[ \frac{ \prod_{i=1}^{p-1}(N-n_{i})! }{(N-n_p)! (N!)^{p-2}}\right]^{1/2} \,.
\ee
The same procedure can be applied to obtain the following extremal correlators,
\be
\langle O_{n_{p}}^{(2,2) \dagger} O_{n_{p-1}}^{(2,2)} O_{n_{p-2}}^{(0,0)} \ldots  O_{n_{1}}^{(0,0)} \rangle
&=& F_p(n_i) \frac{ (n_p)^{p-7/2} (n_{p-1})^{3/2}}{( n_{p-2} \cdots n_1 )^{1/2}}\, ,
\\
\langle O_{n_{p}}^{(b,\bar{b}) \dagger} O_{n_{p-1}}^{(a,\bar{a})} O_{n_{p-2}}^{(0,0)} \ldots  O_{n_{1}}^{(0,0)} \rangle
&=& \delta^{ab} \delta^{\bar{a}\bar{b}} F_p(n_i) \frac{(n_p)^{p-5/2} ( n_{p-1})^{1/2}}{(n_{p-2} \cdots n_1 )^{1/2}}\, ,
\\
\langle O_{n_{p}}^{(2,2) \dagger} O_{n_{p-1}}^{(a,\bar{a})} O_{n_{p-2}}^{(b,\bar{b})} O_{n_{p-3}}^{(0,0)} \ldots  O_{n_{1}}^{(0,0)} \rangle
&=& \e^{ab} \e^{\bar{a}\bar{b}} F_p(n_i) \frac{(n_p)^{p-7/2} (n_{p-1} n_{p-2})^{1/2}}{( n_{p-3} \cdots n_1 )^{1/2}}\, .
\label{ferlong}
\ee
The large $N$ limit is obtained  using
\be
\lim_{N \rightarrow \infty} F_p(n_i) = \left( \frac{1}{N} \right)^{\frac{p-2}{2}} \,.
\ee
According to \cite{deBoer:2008ss},  extremal correlators  are not renormalized under marginal deformations away from the orbifold point, so one
expects the same expressions in the string/supergravity dual.

\subsection{Extremal correlators compute Hurwitz numbers}
If we now look back at our results, we notice that the five types of correlators  (four polynomial and one non-polynomial)
can all be expressed in a uniform way by performing the rescaling
\be\label{normExt}
O_{n}^{(0,0)} &\to& {\hat O}_{n}^{(0,0)}=n^{1/2}\,O_{n}^{(0,0)},
\\
 {O}_{n}^{(0,0)\,\dagger}&\to& {\hat O}_{n}^{(0,0)\,\dagger}= n^{-3/2}{O}_{n}^{(0,0)\,\dagger} ,
\\
 O_{n}^{(a,\bar a)} &\to& {\hat O}_{n}^{(a,\bar a)}=n^{-1/2}\,O_{n}^{(a,\bar a)},
\\
O_{n}^{(a,\bar a)\dagger} &\to& {\hat O}_{n}^{(a,\bar a)\dagger}=n^{-1/2}\,O_{n}^{(a,\bar a) \dagger},
\\
O_{n}^{(2,2)} &\to& {\hat O}_{n}^{(2,2)}=n^{-3/2}\,O_{n}^{(2,2)},
\\
O_{n}^{(2,2)\dagger}&\to& {\hat O}_{n}^{(2,2)\dagger}=n^{1/2}\,O_{n}^{(2,2) \dagger}.
\ee
With this rescaling the two-point functions become
\be
\langle{\hat O}_{n}^{(a,\bar a) \, \dagger}{\hat O}_{n}^{(a,\bar a)}\rangle
=\langle{\hat O}_{n}^{(2,\bar 2) \, \dagger}{\hat O}_{n}^{(2,2)}\rangle=\langle{\hat O}_{n}^{(0,0) \, \dagger}{\hat O}_{n}^{(0,0)}\rangle=\frac{1}{n},
\ee
and the five structure constants of the chiral ring become, at large $N$,\footnote{
As in \eqref{nfer1} and \eqref{nfer2}, for
 the two types of extremal correlators with $a,\bar{a}$ indices we have to add appropriately $\delta$s
and $\e$s to \eqref{tpco}, \eqref{hurnumcorr}, and \eqref{finhur}.}
\be
\langle {\hat O}_{n_3}^{\dagger} {\hat O}_{n_2} {\hat O}_{n_1} \rangle = \left( \frac{1}{N} \right)^{1/2}.
\label{tpco}
\ee
Remarkably, the five types of  extremal correlators are now given by the simple expression
\be\label{hurnumcorr}
\hat C_p = F_p(n_i)\, H_p(\{n_i\}) \ ,
\ee where $H_p(\{n_i\})$ is, as defined above \eqref{hurext} for the polynomial cases, the number of maps contributing to a given correlator.
The relation \eqref{hurnumcorr}  also holds, at large $N$,  for the non-polynomial four-point function \eqref{extraextC},
with $H_4(\{n_i\})$ given in~(\ref{dnonpol}).

The elegance of this result suggests that this relation might hold also for more general extremal correlators, which include (properly rescaled) multi-cycle states. In this case, the Hurwitz numbers count the number of maps with multi-cycle branching points.
More precisely, we conjecture that in the  large $N$ limit the non-vanishing extremal correlators satisfy
\be\label{finhur}
\langle{\hat O}_{[g_0]}^{{\mathcal A}_0\,\dagger}(z_0,\bar z_0)\prod_{i=1}^{p-1} {\hat O}_{[g_i]}^{{\mathcal A}_i}(z_i,\bar z_i)\rangle
= \frac{1}{N^{\frac{p}{2}-1}}\, H_p(\{[g_i]\}),
\ee where $H_p(\{[g_i]\})$ is the number of maps from the base sphere to the covering sphere with $p$ ramifications of type $[g_i]$ at points $z_i$,
and ${\mathcal A}_i$ denote the additional quantum numbers. Since  the Hurwitz numbers are topological invariants
it is likely that it will be possible to prove our conjecture  from first principles by performing a topological twist of the symmetric product theory,
and using localization techniques.\footnote{See e.g.~\cite{Bouchard:2007hi} for a relation between Hurwitz numbers
and topological strings.} The topological A and B models, obtained in the standard fashion by twisting the $(2,2)$ supersymmetry,
are equivalent in this case, because the model has $(4,4)$ supersymmetry. It is plausible that the full power of this bigger symmetry
may play a role in proving our conjecture  and extending our results, much as it did in the proof of the non-renormalization theorem~\cite{deBoer:2008ss}.
The investigation of these ideas is left for future work.

\vskip 1.5 cm

\section*{Acknowledgements}
We thank Antal Jevicki, Samir Mathur, Amit Sever and Cristian Vergu for discussions,
and Congkao Wen for collaboration at the early stages of this work.
SSR would like to thank the HET group at the Weizmann Institute for hospitality during final stages
of this project. LR would like to thank the KITP, Santa Barbara and the Galileo Galilei Institute, Florence,
for hospitality during the completion of this work.
The work of AP was supported in part by DOE grant DE-FG02-91ER40688 and
NSF grant PHY-0643150. The work of LR and SSR is supported in part by DOE
grant  DEFG-0292-ER40697
and by  NSF grant PHY-0653351-001. Any opinions, findings, and conclusions or recommendations expressed in this material are those of the authors and do not necessarily reflect  the views of the National Science Foundation.

\newpage

\appendix

\section{Details of the polynomial map}\label{details}

In this Appendix we derive the map relevant for the correlators
discussed in Section \ref{mapsec}.

We discuss a map from a covering sphere, $t$, to a base sphere, $z$,
with ramification points of order $n$ at $z=0$, order $2$ at $z=1$,
order $n+2$ at $z=\infty$ and order $2$ at $z=u$.
We take the images of the ramification points to be at $t=0,1,\infty$
and $t=x$ respectively. The relation between $x$ and $u$ will
be derived shortly.
The derivative of the map is given by
\be
z'(y)&=&C\, (y+x)^{n-1}\,(y+x-1)\, y,
\ee where $y=t-x$. Integrating the above we get
\be
z(y)=%(x + y)^n \frac{-n x^2 + n y (-2 + n (-1 + y) + y) +
  %x (2 + n + n^2 y)}{n (2 + 3 n + n^2)}+u(x).\nonumber
C\,(x + y)^n \frac{n(n+1)y^2+(n^2(x-1)-2n)y+x(2+n-nx)}{n (2 + 3 n +
n^2)}+v(x).
\ee
We set $C$ and $v(x)$ by demanding $z(y=1-x)=1$ and  $z(y=-x)=0$,
\be
v(x)=x^{1 + n} \frac{2 + n - n x}{(n+2) x-n },\quad
z(t)=t^n \frac{n(n+1)\,t^2-n(n+2)(1+x)\,t+(n+2)(n+1)x}{(n+2)
x-n}.\nonumber\\
\ee
The relation between $u$ and $x$ is set by demanding $x$ to satisfy $u=v(x)$.

Explicit construction of the differential equation satisfied by a four-point function requires
computing several quantities built from the map in the limit of $t
\to x$ \eqref{gabcd}, e.g. the Schwarzian derivative.
 Lets us find the expansion of $z-u$ in terms of
$y$. Writing $z-u=y^2\sum_k a_k y^k$ we obtain for the first
coefficients,
\be
a_0=
\frac{1}
{\sum_{k=0}^{n-1}\sum_{l=0}^{1}\frac{2}{k+l+2} (-1)^{k+l+2}
\left(
\begin{array}{c}
n-1  \\
k
\end{array}
\right)
\left[x^{-k}(x-1)^{2+k}-x^{2+l}(x-1)^{-l}\right]\,
},
\ee
\be
a_1=
\frac{2}{3}a_0\,\left(\frac{n-1}{x}+\frac{1}{x-1}\right),\qquad
a_2=
\frac{1}{2}a_0\,\left(\frac{n-1}{x(x-1)}+
\frac{(n-1)(n-2)}{2x^2}\right).
\ee
We can also compute the inverse expansion,
\be
y=\sum_{k=1}^\infty c_k(z-u)^{k/2}.
\ee The different expansion coefficients are related as
\be
c_1=a_0^{-1/2},\qquad c_2=-\frac{a_1a_0^{-2}}{2},\qquad
c_3=\frac{a_0^{-3/2}}{8}\left[5\frac{a_1^2}{a_0^{2}}-4
\frac{a_2}{a_0}\right].
\ee

The above results are needed to write down the differential equation
(\ref{gabcd}).
We get for the quantities appearing in this equation,
\be
\left(\frac{t''}{t'}\right)'&=&\frac{1}{2}\frac{1}{(z-u)^2}-
\frac{c_2}{c_1}\frac{1}{2}(z-u)^{-3/2}+\dots\\
\left(\frac{t''}{t'}\right)^2&=&\frac{1}{4}\frac{1}{(z-u)^2}-
\frac{c_2}{c_1}(z-u)^{-3/2}+3
\left[\frac{c_2^2}{c_1^2}-\frac{c_3}{c_1}\right](z-u)^{-1}+\dots
\nonumber
\ee From here we obtain
\be
\left(\frac{t''}{t'}\right)'-\half \left(\frac{t''}{t'}\right)^2&=&
\frac{3}{8}\frac{1}{(z-u)^2}-
\frac{3}{2}\left[\frac{c_2^2}{c_1^2}-\frac{c_3}{c_1}\right](z-u)^{-1}+\dots .
\ee We will also need the following
\be
(t')^2&=&
\frac{c_1^2}{4}(z-u)^{-1}+c_1c_2(z-u)^{-1/2}+\frac{1}{2}\left(2c_2^2
+3c_1c_3\right)\dots
\nonumber\\
\frac{1}{t-x}&=&
\frac{1}{c_1}(z-u)^{-1/2}-\frac{c_2}{c_1^2}+\frac{c_2^2-c_1c_3}{c_1^3}(z-u)^{1/2}+\dots\\
\frac{1}{t-x+a}&=&
\frac{1}{a}-\frac{c_1}{a^2}(z-u)^{1/2}-\left[\frac{c_2}{a^2}-\frac{c_1^2}{a^3}\right](z-u)+
\frac{-c_1^3 + 2 a c_1 c_2 - a^2 c_3}{a^4}(z-u)^{3/2}+\dots\;,\nonumber
\ee where $a$ is some complex number.

\

Plugging the above results into \eqref{gabcd}, and integrating the
differential equation, expression \eqref{gform} is obtained.

\section{Counting maps and OPE limits.}

In this appendix we compute the total number of maps, and the number of maps contributing to the $OPE$ limit
in several cases relevant to the discussion in the bulk of the paper. It can be shown that counting covering
maps is equivalent to counting certain types of graphs, and in fact there are many ways to define such graphs~\cite{Lando:2003gx}.
In \cite{PRR1} we developed a diagrammatic language suitable for symmetric product orbifolds.
In what follows we use this language to solve the enumerative problems at hand.
For notations and explanations of the diagrams we refer the reader 
  to~\cite{PRR1}, where   the case of four-point polynomial maps is treated in detail.
 
\subsection{Diagrammatic counting of maps for five-point polynomial correlators.}
\label{ntilde}
\begin{figure}[htbp]
{\tiny
\begin{center}
$\begin{array}{c@{\hspace{0.15in}}c@{\hspace{0.35in}}c}
 \epsfig{file=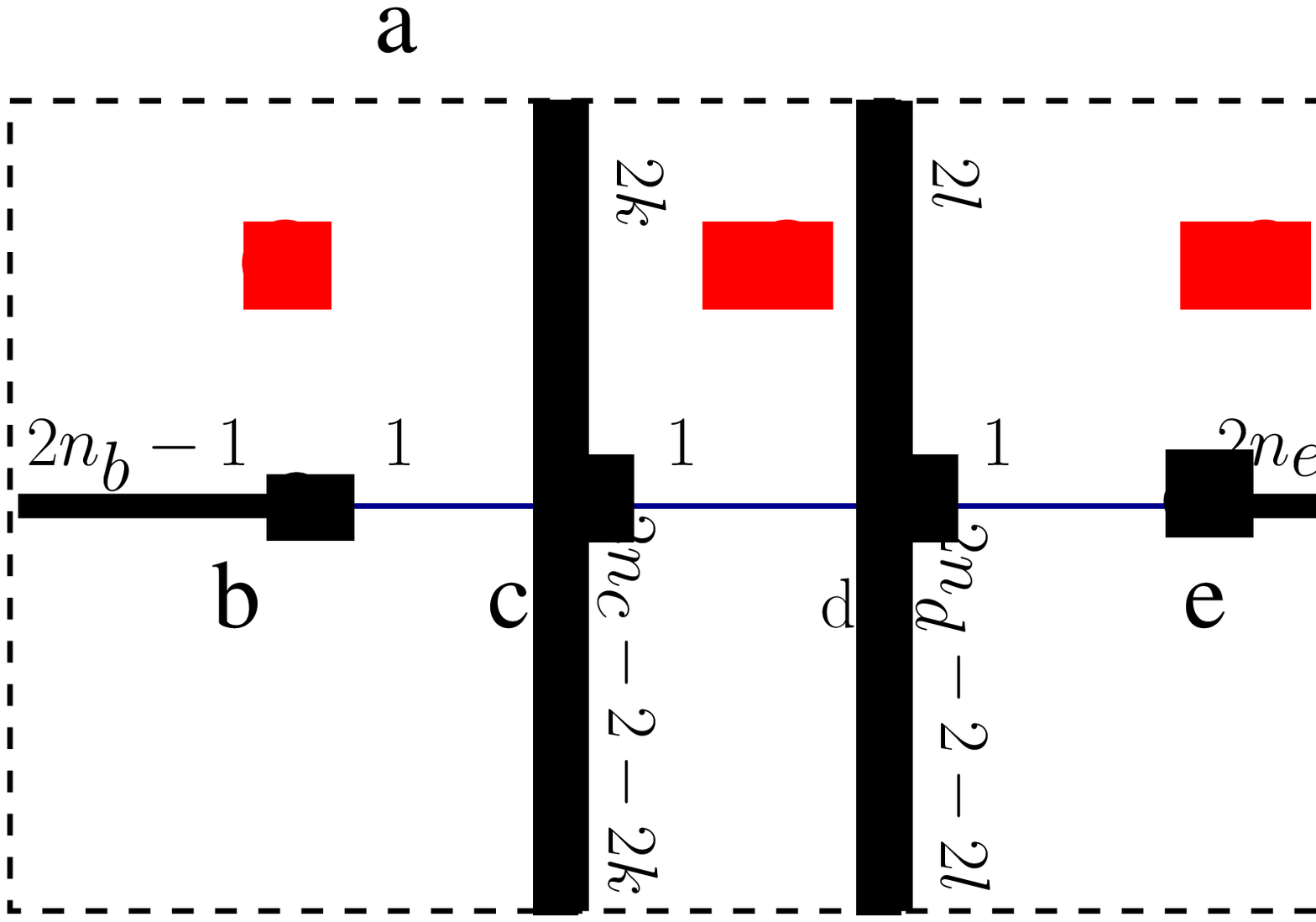,scale=0.2} & \epsfig{file=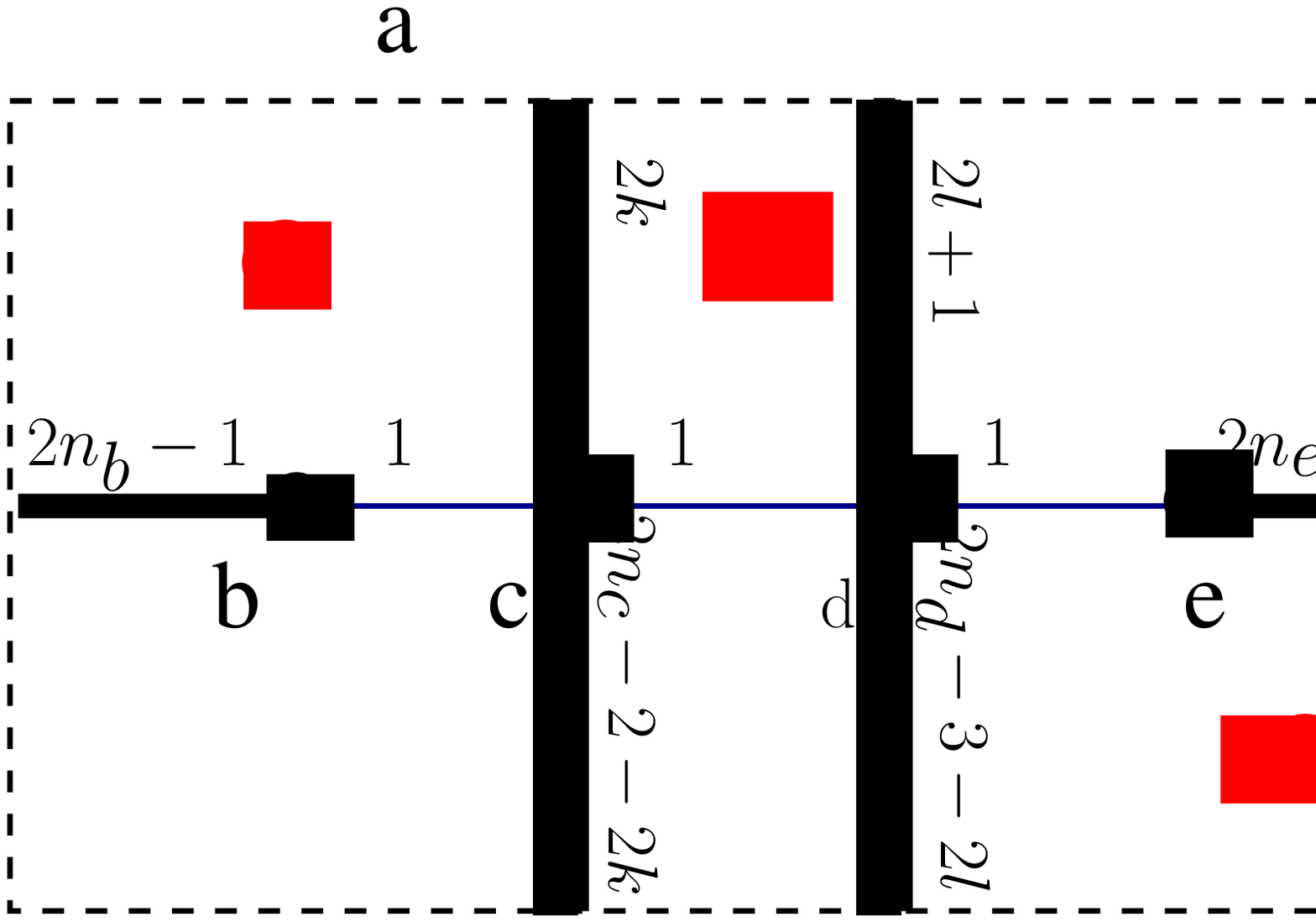,scale=0.2}&\epsfig{file=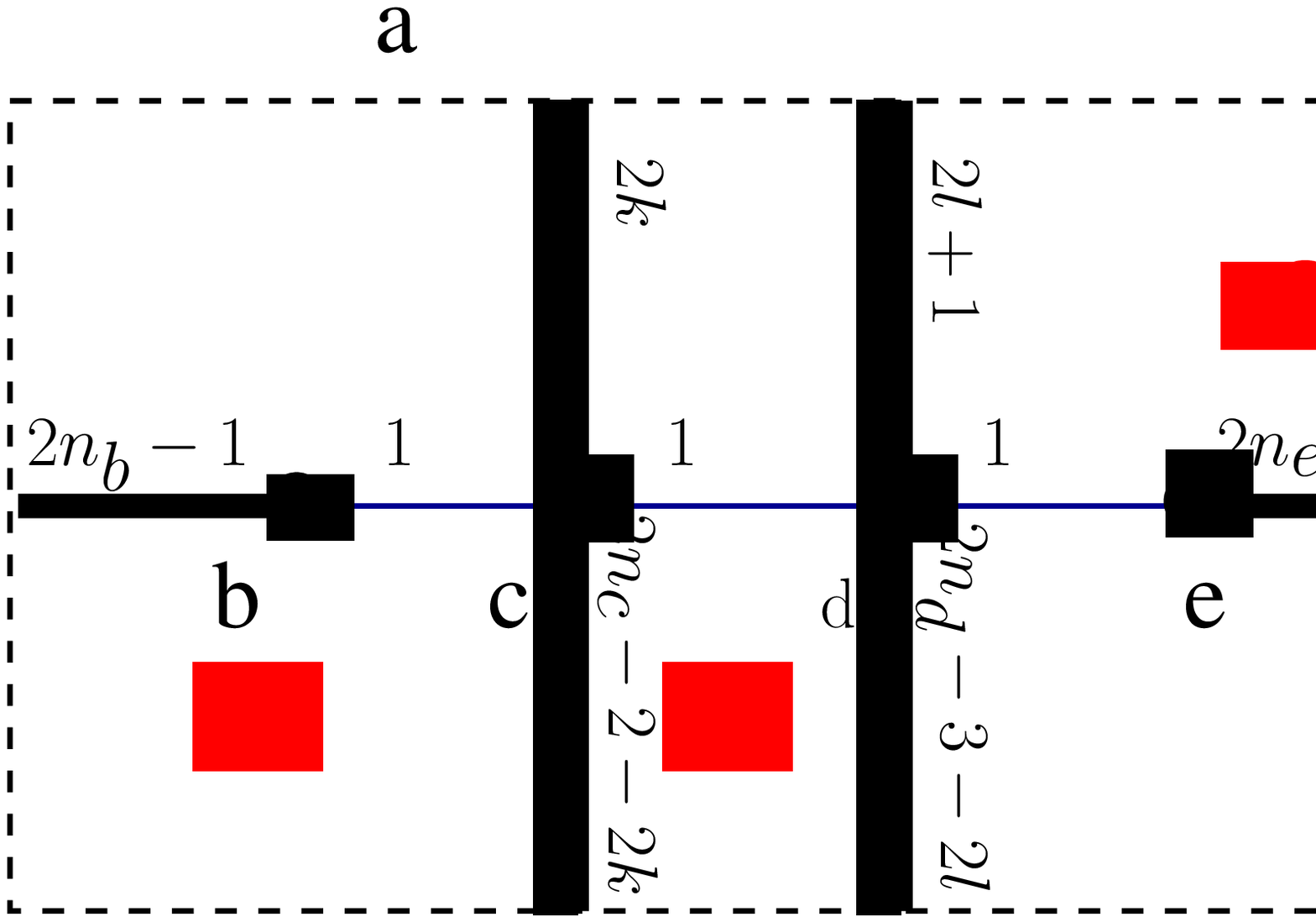,scale=0.2} \\
(I)\;:\;b\;c\;d\;e\;a &(II)\;:\; (e)\,b\;(e)\;c\;(e)\;\;d\;a &(III)\;:\;d\;(e)\;c\;(e)\;\;b\;(e)\;a
\\
\epsfig{file=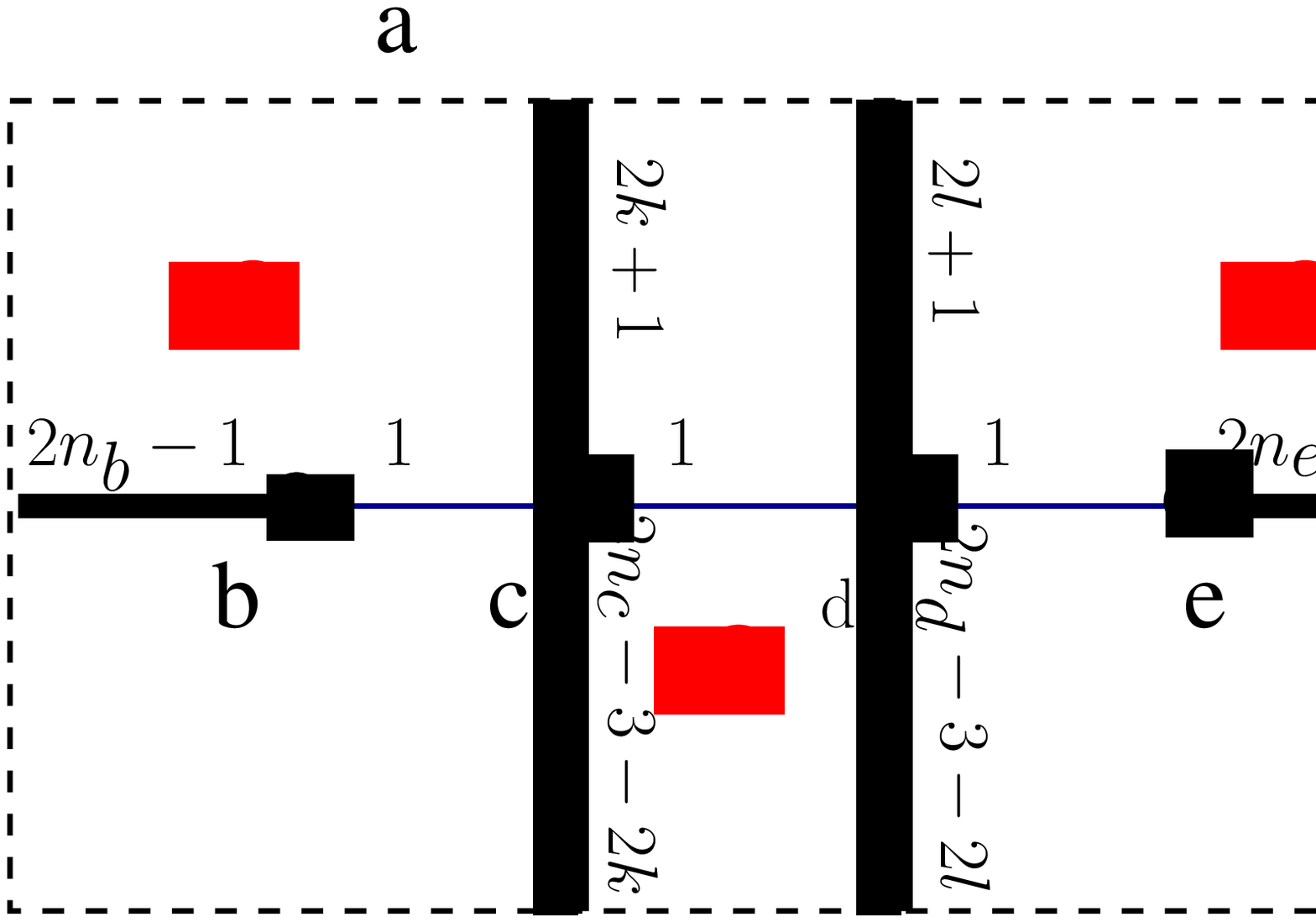,scale=0.2}&\epsfig{file=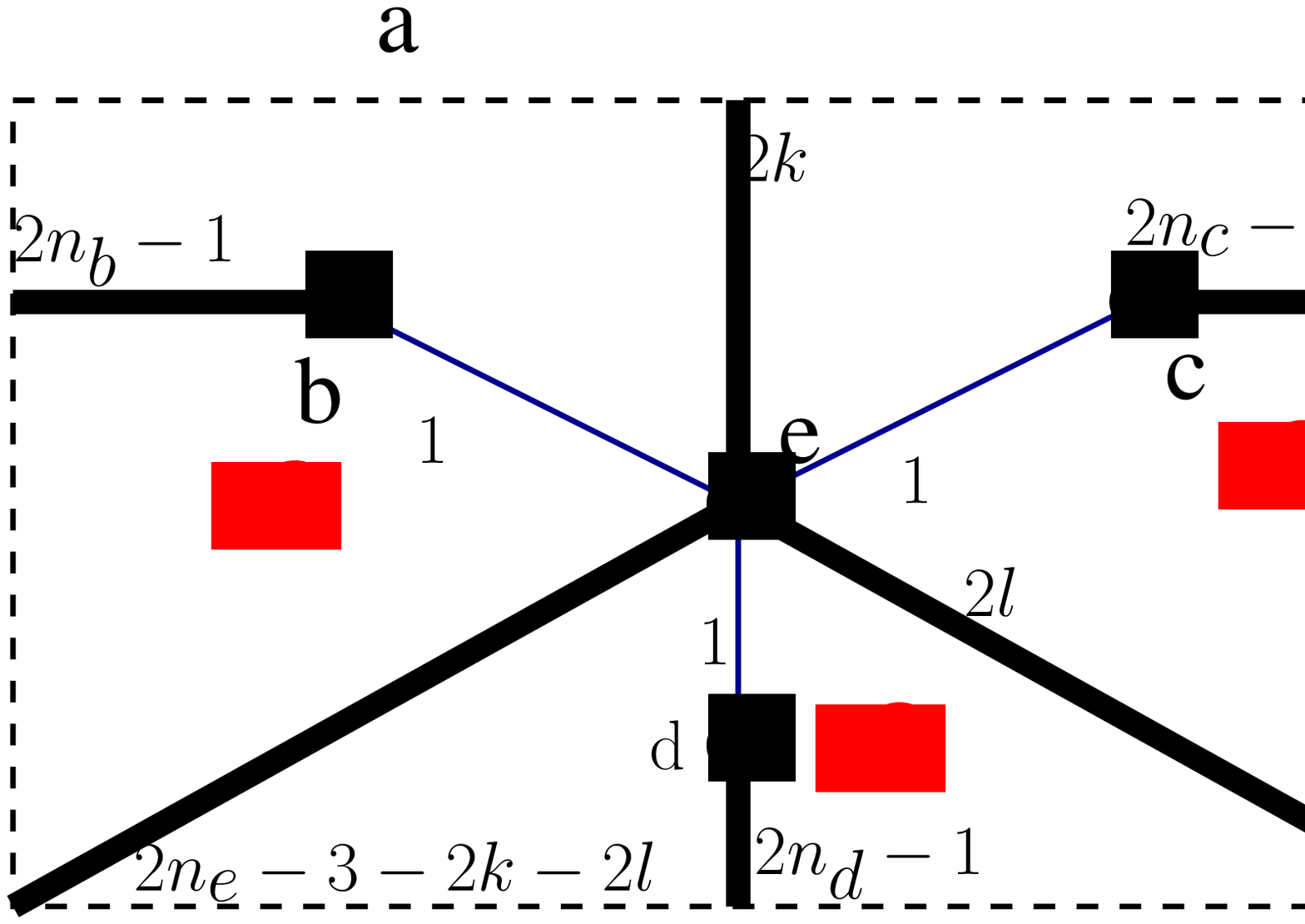,scale=0.2} & \epsfig{file=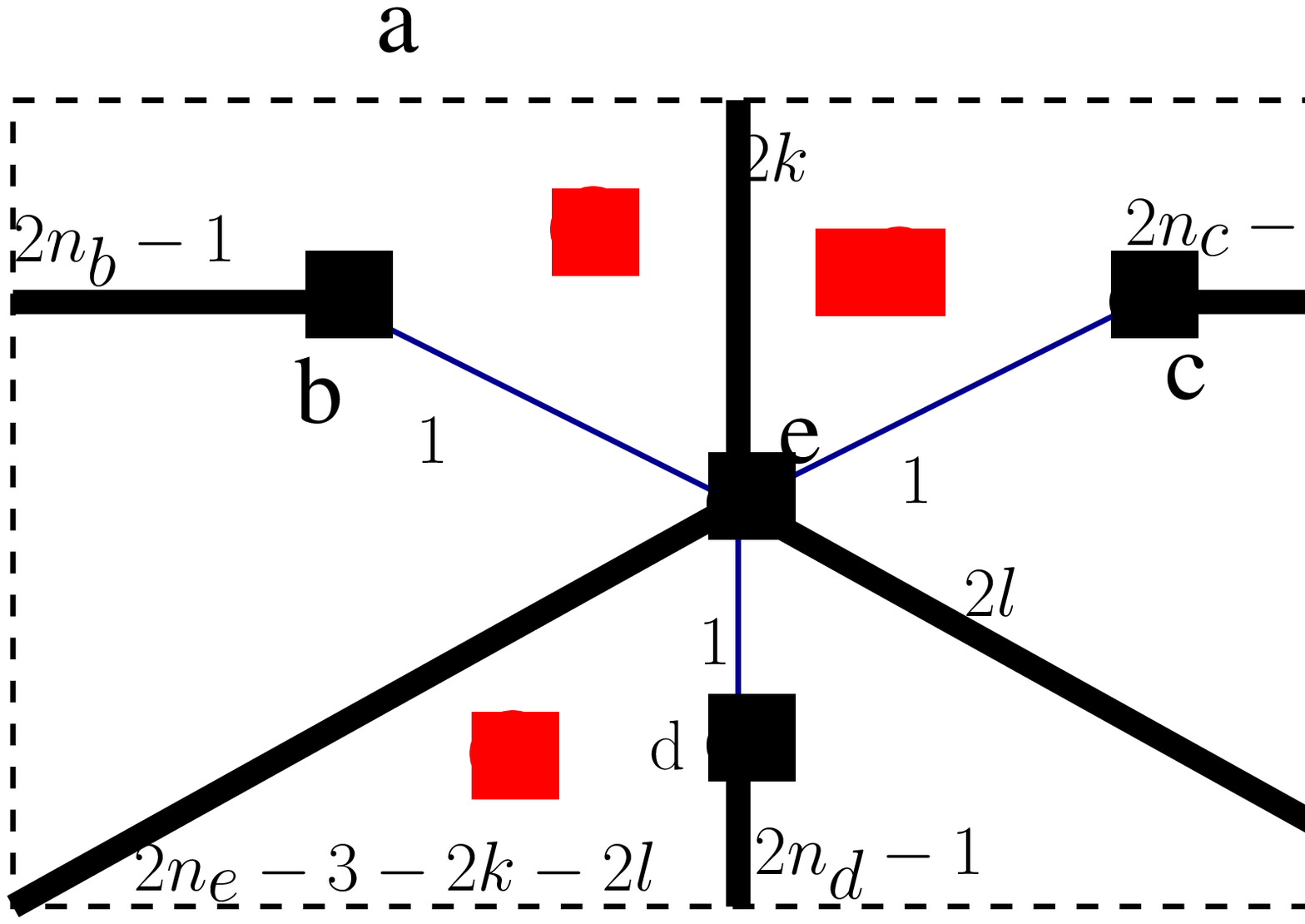,scale=0.2}\\
(IV)\;:\; b\;d\;(e)\;c\;(e)\;a,\;\;d\;(e)\;b\;(e)\;c\;(e)\;a &(V)\;:\;c\;e\;(d)\;b\;(d)\;a &(VI)\;:\; (b)\,d\;(b)\;e\;\;c\;a\\
\epsfig{file=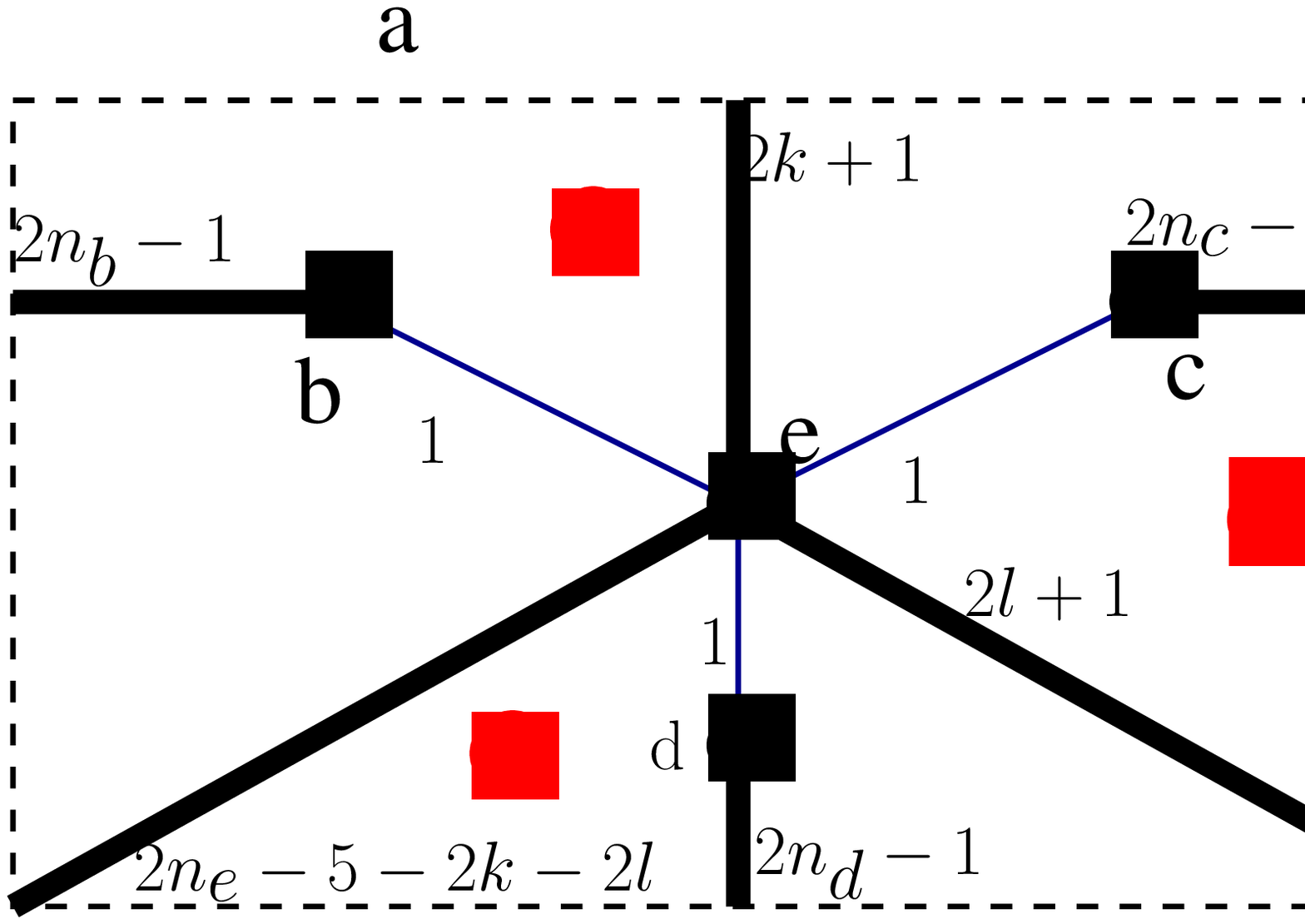,scale=0.2} & \epsfig{file=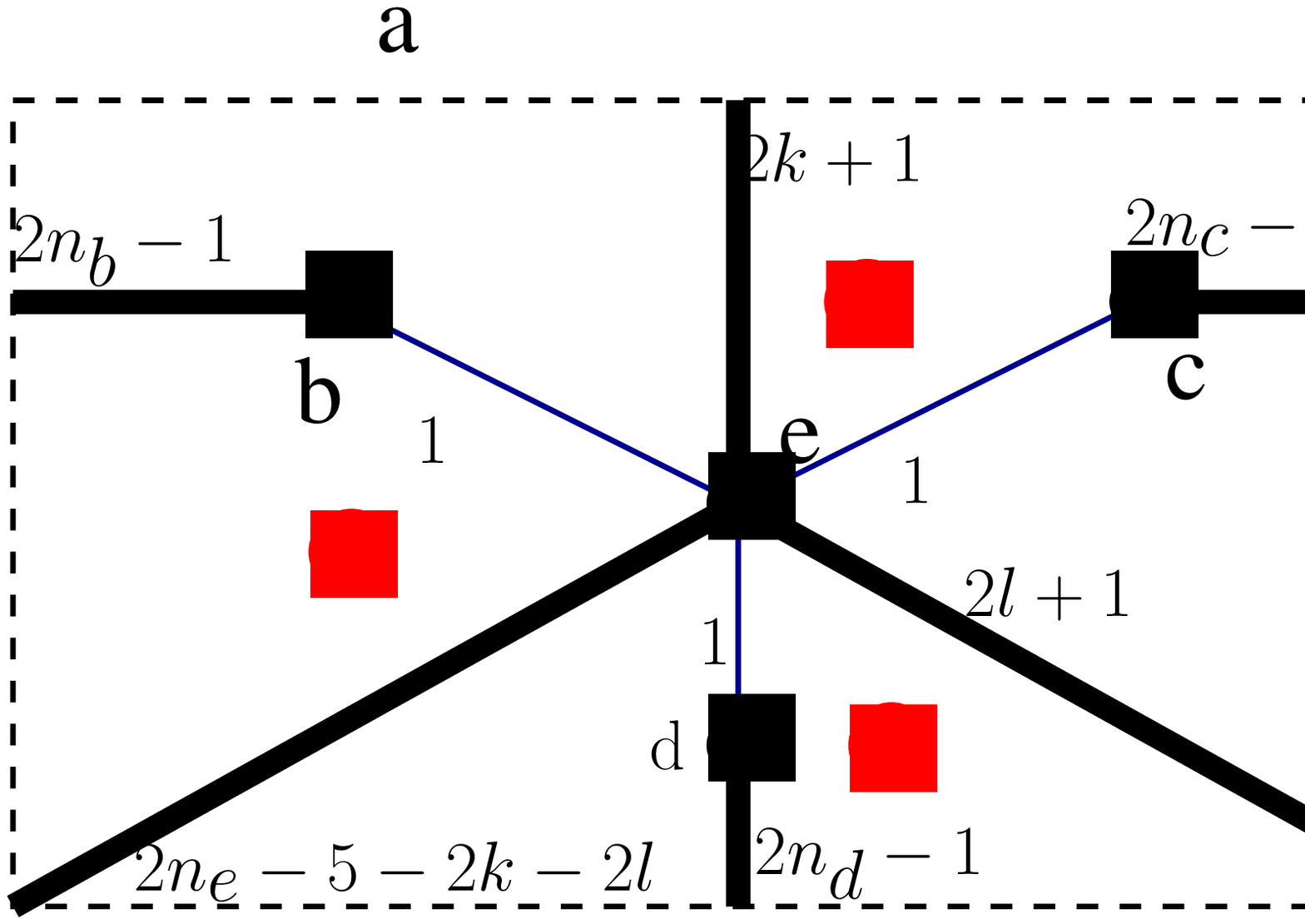,scale=0.2}&\\
 (VII)\;:\;\left[b,\;c,\;d\right]\;e\;a &(VIII)\;:\; e\;\left[b,\;c,\;d\right]\;a&
\\ [0.2cm]
\end{array}$
\end{center}}
 \begin{center}
\caption{The eight classes of different diagrams contributing to a generic polynomial five-point function. The number over each line is the number
of propagators joined in that line. The four vertices at finite positions are $b,c,d,e $, and $a$ is the vertex at infinity. Below each diagram
 we indicate the ordering of the vertices. The commutator denotes that vertices commute, and parenthesis indicate the possible position of an
 operator.
 } \label{5p1}
\end{center}
\end{figure}

We want to compute the total number of maps, and the number of maps contributing to the $OPE$ limit,
in a five-point polynomial correlator.
Consider a  polynomial correlator with cycles $\tilde n_1\leq \tilde n_2\leq \tilde n_3\leq \tilde n_4$
inserted at finite points and cycle $\tilde n_5=\tilde n_1+\tilde n_2+\tilde n_3+\tilde n_4-3$
inserted at infinity.
The different diagrams contributing to the polynomial correlator can be split into eight classes.
These are depicted in Figure \ref{5p1}.
To count the number of diagrams we sum the contributions from each graph and each possible ordering, i.e. each assignment
of the cycle lengths $\tilde n_i$ to the lengths $n_i$ appearing in Figure \ref{5p1}. Essentially  we have to count all
the possible ways to choose the numbers $l$ and $k$ in these diagrams. The results are summarized in  Table~\ref{table5p}.
\vskip .5cm
%\ref{table5p}.
\be\label{table5p}
\begin{tabular}{|c|c|c|c|}
\hline
Class & Ordering & \#\\
\hline\hline
$(I)$&$bcdea$ & $n_2n_3$\\\hline
$(II)$&$ebcda$ & $n_3(n_4-1)$\\\hline
$(II)$&$becda$ & $n_3(n_4-1)$\\\hline
$(II)$&$bceda$ & $n_2(n_4-1)$\\\hline
$(III)$&$decba$ & $(n_1-1)n_3$\\\hline
$(III)$&$dceba$ & $(n_1-1)n_2$\\\hline
$(III)$&$dcbea$ & $(n_1-1)n_2$\\\hline
$(IV)$&$bdeca$ & $(n_2-1)(n_4-1)$\\\hline
\end{tabular}\;
\begin{tabular}{|c|c|c|c|}
\hline
Class & Ordering & \#\\
\hline\hline
$(IV)$&$bdcea$ & $(n_2-1)(n_3-1)$\\\hline
$(IV)$&$debca$ & $(n_1-1)(n_4-1)$\\\hline
$(IV)$&$dbeca$ & $(n_1-1)(n_4-1)$\\\hline
$(IV)$&$dbcea$ & $(n_1-1)(n_3-1)$\\\hline
$(V)$&$c\;e\;(d)\;b\;(d)\;a$ & $n_2(n_2-1)$\\\hline
$(VI)$&$(b)\,d\;(b)\;e\;\;c\;a$ & $n_3(n_3-1)$\\\hline
$(VII)$&$\left[b,\;c,\;d\right]\;e\;a$ & $(n_4-2)(n_4-1)$\\\hline
$(VIII)$&$e\;\left[b,\;c,\;d\right]\;a$ & $(n_1-2)(n_1-1)$\\\hline
\end{tabular}\nonumber\\
\nn
\ee
Summing all the contributions from  table \ref{table5p} we get that the number of maps
in the five point extremal case is $\tilde n_5^2=\left(\tilde n_1+\tilde n_2+\tilde n_3+\tilde n_4-3\right)^2$.
 Let us count now the diagrams which contribute in $OPE$ limit
of $\tilde n_1$ cycle colliding with $\tilde n_2$ cycle. The diagrams which contribute are the ones with a propagator stretched between $\tilde n_1$
 and $\tilde n_2$. Thus,
we count those diagram in table \ref{table5p} having a propagator between the two first cycles in column $Ordering$. We get that the number of such maps
is $(\tilde n_1+\tilde n_2-1)\tilde n_5$.

\subsection{OPE counting}
\label{opecount}
The number $H_p$ of terms in (\ref{hpsum}) is the number of $n_p$-sheeted covering maps from $S^2_{cover}$ to  $S^2_{base}$ with $p$ branching points, with branching numbers $n_1,n_2,\ldots n_p$, and satisfying
\be\label{our}
n_p = \sum_{i=1}^{p-1}n_i -p+2.
\label{extremalp2}
\ee The maps in this case are polynomial.
This number can be shown~\cite{Lando2} to be equal to
\be
\label{lando}
H_p=n_p^{p-3}.
\ee
Let us call $\tilde n$ to the the number of maps, out of $H_p$, such that the
OPE of the operator with quantum number $n_1$ and to operator with quantum number $n_2$ gives a single cycle of size $n_1+n_2-1$.

\begin{figure}[htbp]
\begin{center}
 \epsfig{file=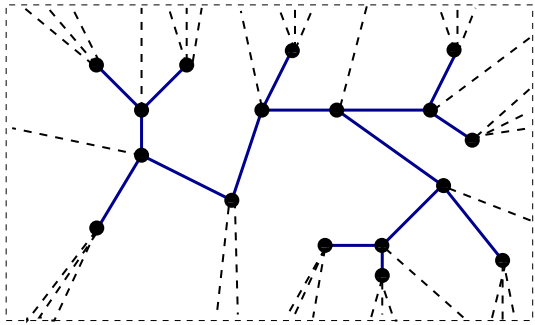,scale=0.4}
\end{center}
 \begin{center}
\caption{An example of a generic graph of a polynomial map. The blue lines connect the first $p-1$ ramification points and the dashed black lines
go to ramification point $n_p$. In this case we have a particular example of a diagram contributing to $\langle(\s_{[2]})^{17}\,\s_{[18]}\rangle $.
 } \label{genext}
\end{center}
\end{figure}

To determine $\tilde n$  consider an auxiliary $(p-1)$-point correlator  with cycles of length $(n_1+n_2-1),\, n_3,\,n_4,\dots,\,n_{p-1},\, n_p$.
From \eqref{lando} there are $n_p^{p-4}$ such maps. To obtain a $p$-point correlator of form \eqref{our} we have
to split the $n_1+n_2-1$ cycle of the auxiliary correlator into $n_1$ and $n_2$ cycle in all possible ways. Obviously, there are $n_1+n_2-1$
ways to do so, so the total number of maps is
\be
\tilde n= (n_1+n_2-1)\, n_p^{s-4}.
\ee
Note that in correlators satisfying \eqref{our}, two vertices can either have no common colors or just a single common color. And thus in the OPE between a pair of consecutive vertices, either the cycles $n_i$ and $n_{i+1}$ join to a single cycle of length $n_i+n_{i+1}-1$ or to a double
cycle of the form $(n_i)(n_{i+1})$. It is easy to see this fact diagrammatically. Any diagram  of such $p$-point correlator has all the
propagators, except $p-2$ ones, going to the vertex $n_p$, and the remaining $p-2$ propagators connect the first $p-1$ vertices into some connected tree structure.
This is exemplified in Figure \ref{genext}.
It is easy to see that two vertices either do not have any common loop and thus no common color, or they have a single common color loop.

\subsection{Planar maps in the extremal non-polynomial four-point function}
\label{nonpol}
\begin{figure}[htbp]
\begin{center}
$\begin{array}{c@{\hspace{0.05in}}c@{\hspace{0.05in}}c}
 \epsfig{file=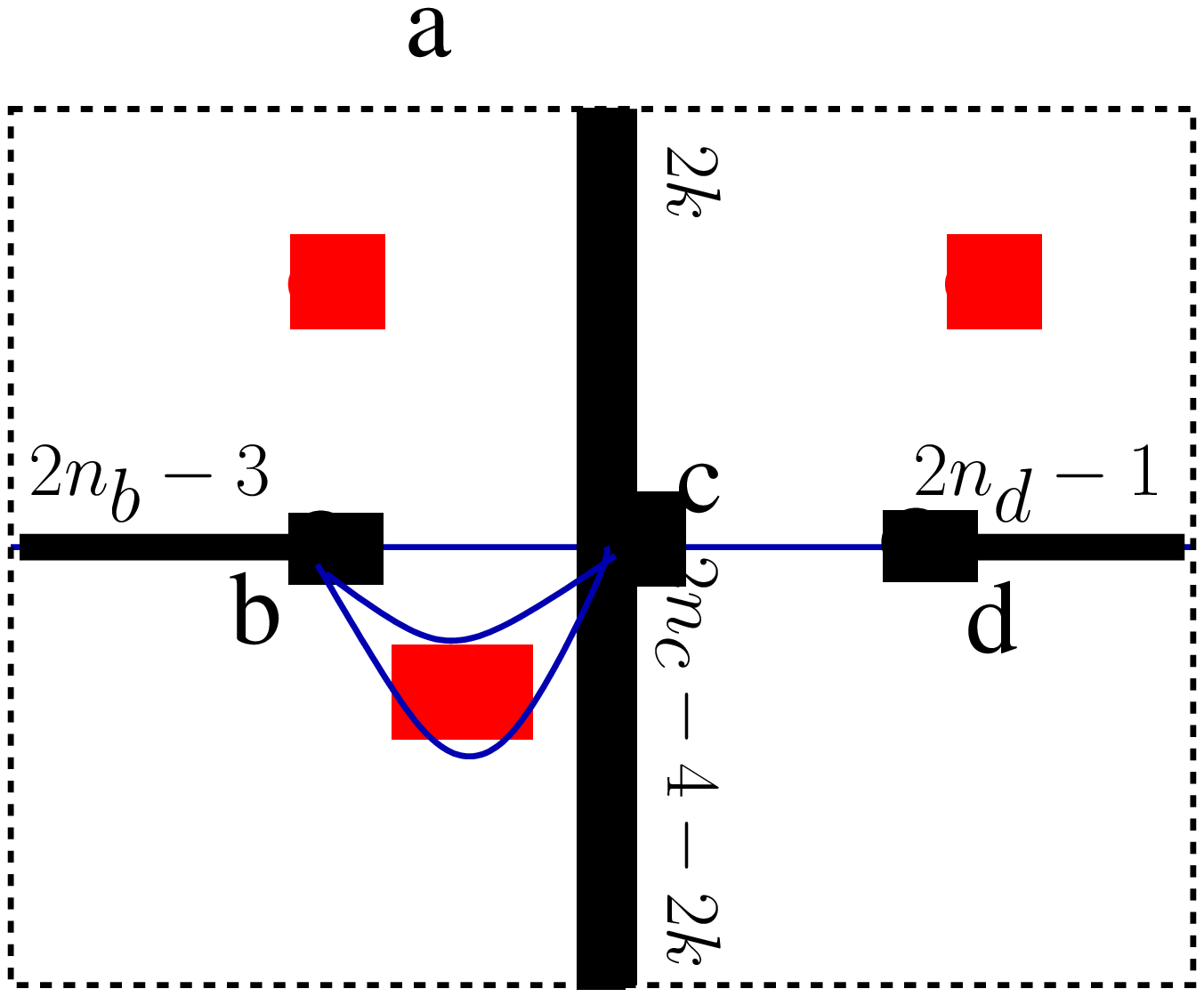,scale=0.3} & \epsfig{file=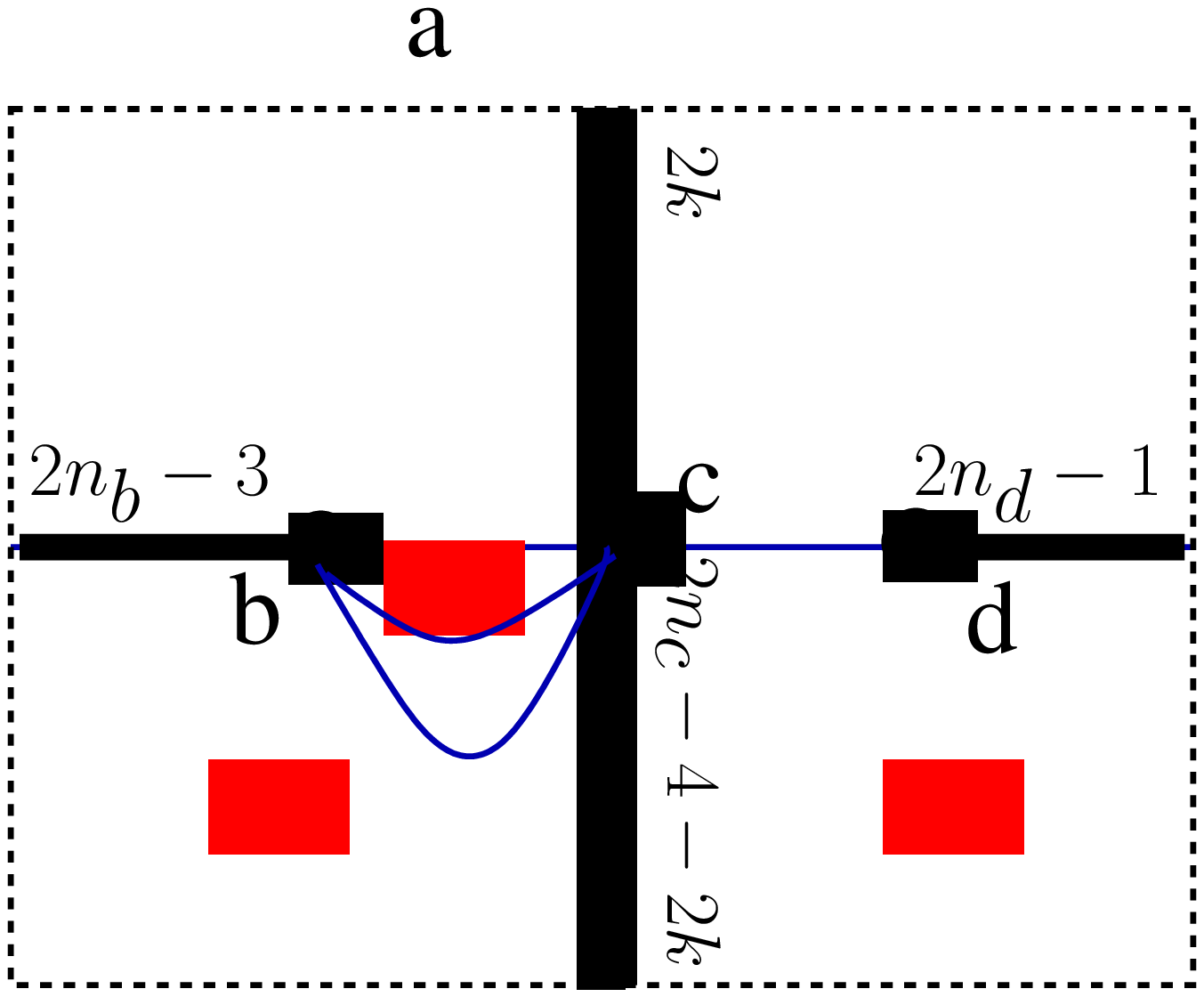,scale=0.3}& \epsfig{file=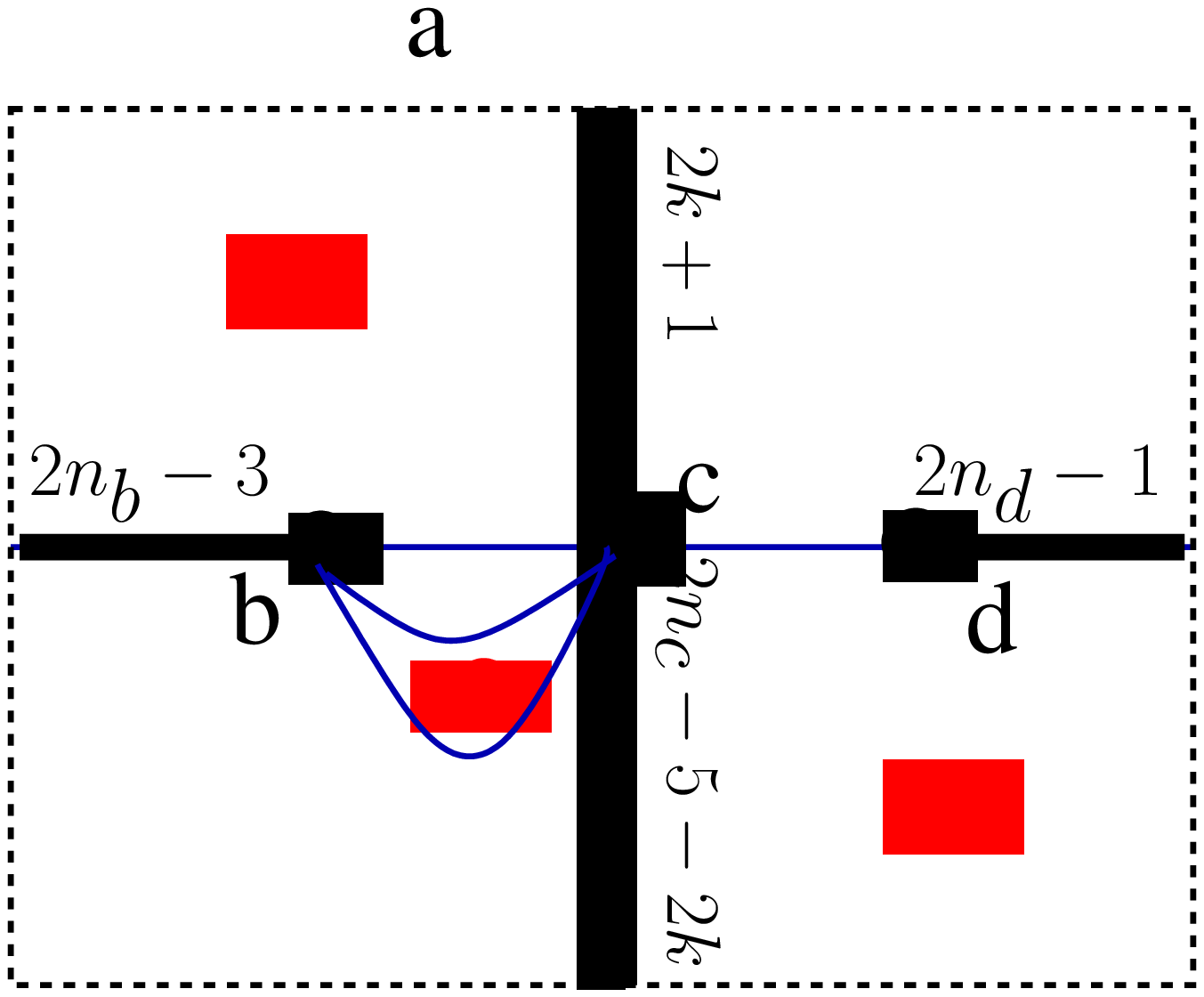,scale=0.3}\\
(I)\;:\;b\;c\;d\;a &(II)\;:\; d\;c\;b\;a& (III)\;:\; [b,\;d]\;c\;a
\\
 \epsfig{file=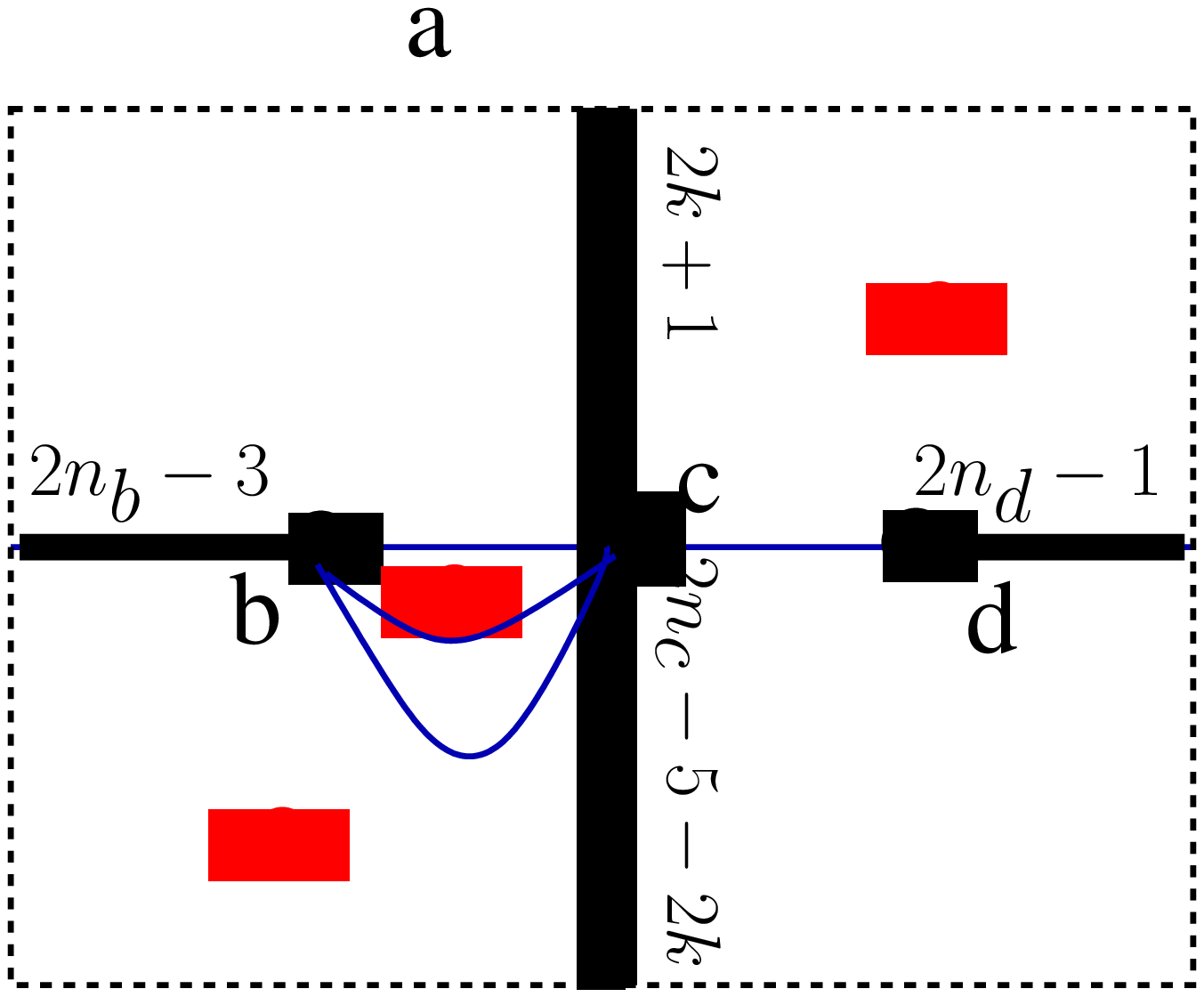,scale=0.3} & \epsfig{file=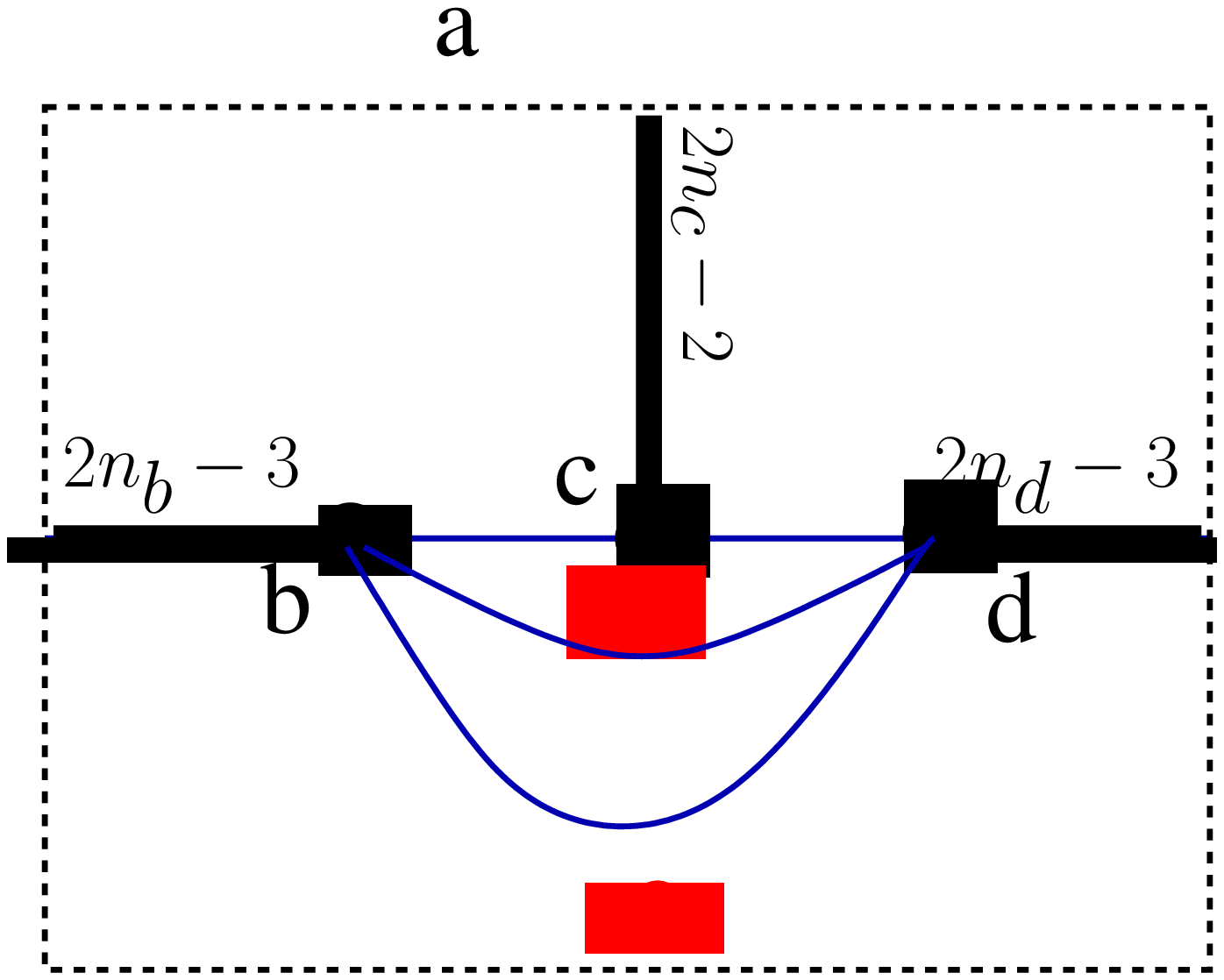,scale=0.3}& \epsfig{file=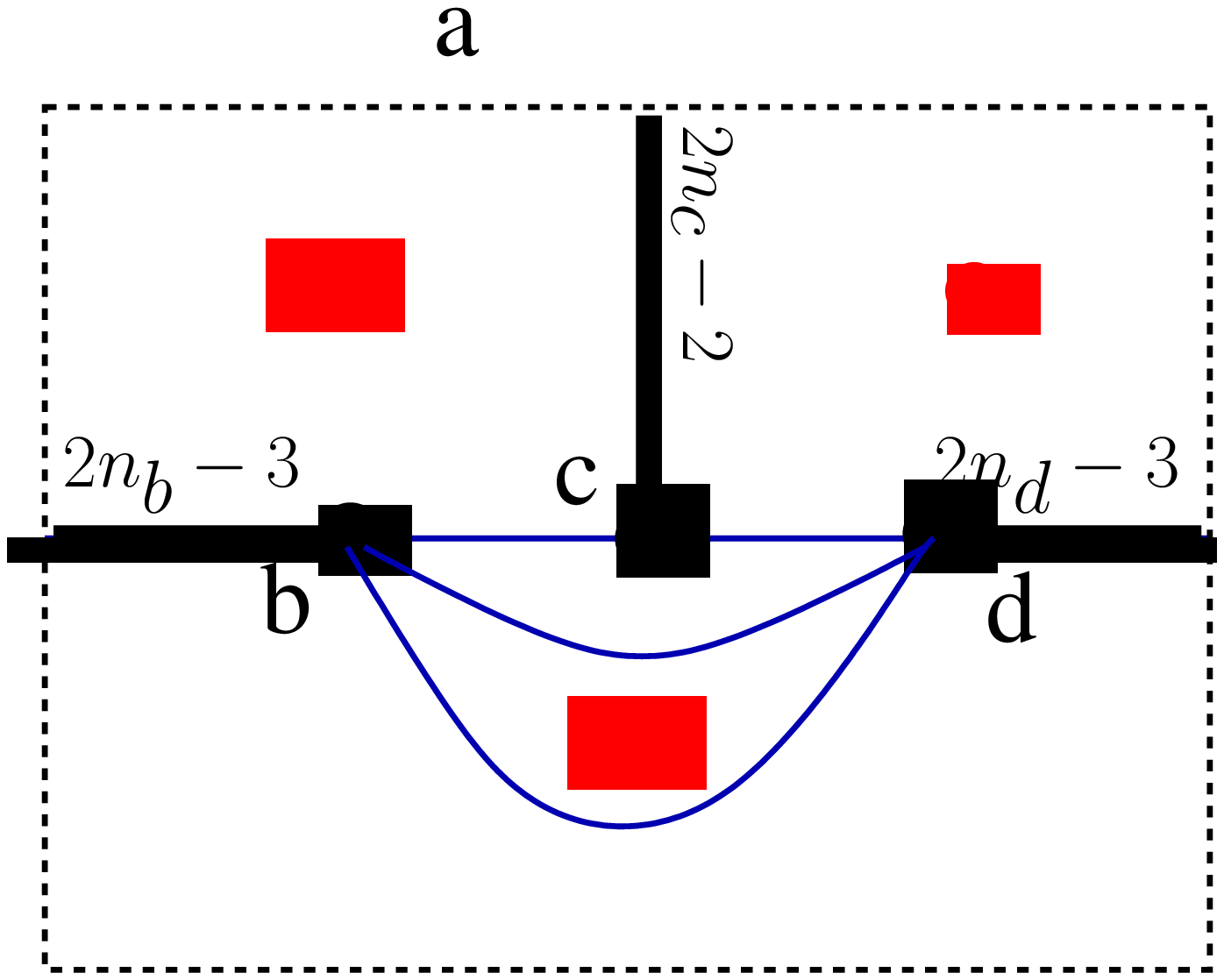,scale=0.3}\\
 (IV)\;:\; c\;[b,\;d]\;a &(V)\;:\; d\;c\;b\;a & (VI)\;:\; b\;c\;d\;a
\\ [0.2cm]
\end{array}$
\end{center}
 \begin{center}
\caption{The different classes of diagrams contributing to the `near polynomial' four-point function. Thin blue lines
represent a single propagator of the graph.
 } \label{a4p}
\end{center}
\end{figure}

In this section we count diagrammatically all the planar maps contributing to the ``near polynomial''  case, $n_4=n_1+n_2+n_3-4$.
All the classes of diagrams are depicted in Figure~\ref{a4p}.
We can count the number of diagrams in each class by just counting the number of values the parameter $k$ can take in each
each class of diagrams in Figure \ref{a4p}. The results are summarized in table \ref{tablea4p}. The column labeled {\it OPE}
denotes whether the diagrams contribute in the OPE limit $(i)$ when two contracted cycles have two indices in common,
or $(ii)$ when a single index is common.
\be\label{tablea4p}
\begin{tabular}{|c|c|c|c|}
\hline
Class & Ordering & \# & OPE\\
\hline\hline
$(I)$&$bcda$ & $n_2-1$&$i$\\\hline
$(II)$&$dcba$ & $n_2-1$&$ii$\\\hline
$(III)$&$bdca$ & $n_3-2$&$-$\\\hline
$(III)$&$dbca$ & $n_3-2$&$-$\\\hline
$(IV)$&$cbda$ & $n_1-2$&$i$\\\hline
$(IV)$&$cdba$ & $n_1-2$&$ii$\\\hline
$(V)$&$dcba$ & $1$&$ii$\\\hline
$(VI)$&$bcda$ & $1$&$ii$\\\hline
\end{tabular}\;
\ee
Summing all the contributions we get that the number of maps is $2(n_1+n_2+n_3-4)=2\,n_4$. The number
of maps contributing to $(i)$ OPE limit is $n_1+n_2-3$, and to the $(ii)$ OPE limit is $n_1+n_2-1$.

\newpage
\bibliography{h3bib}

\providecommand{\href}[2]{#2}\begingroup\raggedright\begin{thebibliography}{10}

\bibitem{PRR1}
A.~Pakman, L.~Rastelli, and S.~S. Razamat, {\it {Diagrams for Symmetric
  Orbifolds}},  \href{http://xxx.lanl.gov/abs/0905.3448}{{\tt 0905.3448}}.

\bibitem{Maldacena:1997re}
J.~M. Maldacena, {\it The large n limit of superconformal field theories and
  supergravity},  {\em Adv. Theor. Math. Phys.} {\bf 2} (1998) 231--252,
  [\href{http://xxx.lanl.gov/abs/hep-th/9711200}{{\tt hep-th/9711200}}].

\bibitem{Aharony:1999ti}
O.~Aharony, S.~S. Gubser, J.~M. Maldacena, H.~Ooguri, and Y.~Oz, {\it Large n
  field theories, string theory and gravity},  {\em Phys. Rept.} {\bf 323}
  (2000) 183--386, [\href{http://xxx.lanl.gov/abs/hep-th/9905111}{{\tt
  hep-th/9905111}}].

\bibitem{Dijkgraaf:2000vr}
R.~Dijkgraaf, {\it On the d1-d5 conformal field theory},  {\em Class. Quant.
  Grav.} {\bf 17} (2000) 1035--1048.

\bibitem{David:2002wn}
J.~R. David, G.~Mandal, and S.~R. Wadia, {\it Microscopic formulation of black
  holes in string theory},  {\em Phys. Rept.} {\bf 369} (2002) 549--686,
  [\href{http://xxx.lanl.gov/abs/hep-th/0203048}{{\tt hep-th/0203048}}].

\bibitem{Martinec}
E.~Martinec, {\it The d1-d5 system},
  \href{http://xxx.lanl.gov/abs/http://hamilton.uchicago.edu/$\sim$ejm/japan99%
.ps}{{\tt http://hamilton.uchicago.edu/$\sim$ejm/japan99.ps}}.

\bibitem{Dijkgraaf:1998gf}
R.~Dijkgraaf, {\it Instanton strings and hyperkaehler geometry},  {\em Nucl.
  Phys.} {\bf B543} (1999) 545--571,
  [\href{http://xxx.lanl.gov/abs/hep-th/9810210}{{\tt hep-th/9810210}}].

\bibitem{Larsen:1999uk}
F.~Larsen and E.~J. Martinec, {\it U(1) charges and moduli in the d1-d5
  system},  {\em JHEP} {\bf 06} (1999) 019,
  [\href{http://xxx.lanl.gov/abs/hep-th/9905064}{{\tt hep-th/9905064}}].

\bibitem{Maldacena:1998bw}
J.~M. Maldacena and A.~Strominger, {\it Ads(3) black holes and a stringy
  exclusion principle},  {\em JHEP} {\bf 12} (1998) 005,
  [\href{http://xxx.lanl.gov/abs/hep-th/9804085}{{\tt hep-th/9804085}}].

\bibitem{deBoer:1998ip}
J.~de~Boer, {\it Six-dimensional supergravity on s**3 x ads(3) and 2d conformal
  field theory},  {\em Nucl. Phys.} {\bf B548} (1999) 139--166,
  [\href{http://xxx.lanl.gov/abs/hep-th/9806104}{{\tt hep-th/9806104}}].

\bibitem{Kutasov:1998zh}
D.~Kutasov, F.~Larsen, and R.~G. Leigh, {\it String theory in magnetic monopole
  backgrounds},  {\em Nucl. Phys.} {\bf B550} (1999) 183--213,
  [\href{http://xxx.lanl.gov/abs/hep-th/9812027}{{\tt hep-th/9812027}}].

\bibitem{Argurio:2000tb}
R.~Argurio, A.~Giveon, and A.~Shomer, {\it Superstrings on ads(3) and symmetric
  products},  {\em JHEP} {\bf 12} (2000) 003,
  [\href{http://xxx.lanl.gov/abs/hep-th/0009242}{{\tt hep-th/0009242}}].

\bibitem{Jevicki:1998bm}
A.~Jevicki, M.~Mihailescu, and S.~Ramgoolam, {\it Gravity from cft on s**n(x):
  Symmetries and interactions},  {\em Nucl. Phys.} {\bf B577} (2000) 47--72,
  [\href{http://xxx.lanl.gov/abs/hep-th/9907144}{{\tt hep-th/9907144}}].

\bibitem{Lunin:2000yv}
O.~Lunin and S.~D. Mathur, {\it Correlation functions for m(n)/s(n) orbifolds},
   {\em Commun. Math. Phys.} {\bf 219} (2001) 399--442,
  [\href{http://xxx.lanl.gov/abs/hep-th/0006196}{{\tt hep-th/0006196}}].

\bibitem{Lunin:2001pw}
O.~Lunin and S.~D. Mathur, {\it Three-point functions for m(n)/s(n) orbifolds
  with n = 4 supersymmetry},  {\em Commun. Math. Phys.} {\bf 227} (2002)
  385--419, [\href{http://xxx.lanl.gov/abs/hep-th/0103169}{{\tt
  hep-th/0103169}}].

\bibitem{Gaberdiel:2007vu}
M.~R. Gaberdiel and I.~Kirsch, {\it Worldsheet correlators in ads(3)/cft(2)},
  {\em JHEP} {\bf 04} (2007) 050,
  [\href{http://xxx.lanl.gov/abs/hep-th/0703001}{{\tt hep-th/0703001}}].

\bibitem{Dabholkar:2007ey}
A.~Dabholkar and A.~Pakman, {\it Exact chiral ring of ads(3)/cft(2)},  {\em
  Adv. Theor. Math. Phys.} {\bf 13} (2009) 409--462,
  [\href{http://xxx.lanl.gov/abs/hep-th/0703022}{{\tt hep-th/0703022}}].

\bibitem{Pakman:2007hn}
A.~Pakman and A.~Sever, {\it Exact n=4 correlators of ads(3)/cft(2)},  {\em
  Phys. Lett.} {\bf B652} (2007) 60--62,
  [\href{http://xxx.lanl.gov/abs/arXiv:0704.3040 [hep-th]}{{\tt arXiv:0704.3040
  [hep-th]}}].

\bibitem{Taylor:2007hs}
M.~Taylor, {\it Matching of correlators in $ads_3/cft_2$},
  \href{http://xxx.lanl.gov/abs/arXiv:0709.1838 [hep-th]}{{\tt arXiv:0709.1838
  [hep-th]}}.

\bibitem{Giribet:2007wp}
G.~Giribet, A.~Pakman, and L.~Rastelli, {\it {Spectral Flow in AdS(3)/CFT(2)}},
   {\em JHEP} {\bf 06} (2008) 013,
  [\href{http://xxx.lanl.gov/abs/0712.3046}{{\tt 0712.3046}}].

\bibitem{Giribet:2008yt}
G.~Giribet and L.~Nicolas, {\it {Comment on three-point function in
  AdS(3)/CFT(2)}},  \href{http://xxx.lanl.gov/abs/0812.2732}{{\tt 0812.2732}}.

\bibitem{Cardona:2009hk}
C.~A. Cardona and C.~A. Nunez, {\it {Three-point functions in superstring
  theory on AdS3xS3xT4}},  \href{http://xxx.lanl.gov/abs/0903.2001}{{\tt
  0903.2001}}.

\bibitem{deBoer:2008ss}
J.~de~Boer, J.~Manschot, K.~Papadodimas, and E.~Verlinde, {\it {The chiral ring
  of AdS3/CFT2 and the attractor mechanism}},
  \href{http://xxx.lanl.gov/abs/0809.0507}{{\tt 0809.0507}}.

\bibitem{Arutyunov:1997gt}
G.~E. Arutyunov and S.~A. Frolov, {\it Virasoro amplitude from the s(n) r**24
  orbifold sigma model},  {\em Theor. Math. Phys.} {\bf 114} (1998) 43--66,
  [\href{http://xxx.lanl.gov/abs/hep-th/9708129}{{\tt hep-th/9708129}}].

\bibitem{hur}
A.~Hurwitz, {\it {Uber die Anzal der Riemann'sche Fl\"achen mit gegebenen
  Verzweigungpunkten}},  {\em Math. Ann.} {\bf 55} (1902) 51.

\bibitem{Rastelli:2005ph}
L.~Rastelli and M.~Wijnholt, {\it Minimal ads(3)},
  \href{http://xxx.lanl.gov/abs/hep-th/0507037}{{\tt hep-th/0507037}}.

\bibitem{Gopakumar:2005fx}
R.~Gopakumar, {\it {From free fields to AdS. III}},  {\em Phys. Rev.} {\bf D72}
  (2005) 066008, [\href{http://xxx.lanl.gov/abs/hep-th/0504229}{{\tt
  hep-th/0504229}}].

\bibitem{Aharony:2006th}
O.~Aharony, Z.~Komargodski, and S.~S. Razamat, {\it {On the worldsheet theories
  of strings dual to free large N gauge theories}},  {\em JHEP} {\bf 05} (2006)
  016, [\href{http://xxx.lanl.gov/abs/hep-th/0602226}{{\tt hep-th/0602226}}].

\bibitem{David:2006qc}
J.~R. David and R.~Gopakumar, {\it {From spacetime to worldsheet: Four point
  correlators}},  {\em JHEP} {\bf 01} (2007) 063,
  [\href{http://xxx.lanl.gov/abs/hep-th/0606078}{{\tt hep-th/0606078}}].

\bibitem{David:2008iz}
J.~R. David, R.~Gopakumar, and A.~Mukhopadhyay, {\it {Worldsheet Properties of
  Extremal Correlators in AdS/CFT}},  {\em JHEP} {\bf 10} (2008) 029,
  [\href{http://xxx.lanl.gov/abs/0807.5027}{{\tt 0807.5027}}].

\bibitem{D'Hoker:1999ea}
E.~D'Hoker, D.~Z. Freedman, S.~D. Mathur, A.~Matusis, and L.~Rastelli, {\it
  Extremal correlators in the ads/cft correspondence},
  \href{http://xxx.lanl.gov/abs/hep-th/9908160}{{\tt hep-th/9908160}}.

\bibitem{Bianchi:1999ie}
M.~Bianchi and S.~Kovacs, {\it {Non-renormalization of extremal correlators in
  N = 4 SYM theory}},  {\em Phys. Lett.} {\bf B468} (1999) 102--110,
  [\href{http://xxx.lanl.gov/abs/hep-th/9910016}{{\tt hep-th/9910016}}].

\bibitem{Eden:1999kw}
B.~Eden, P.~S. Howe, C.~Schubert, E.~Sokatchev, and P.~C. West, {\it {Extremal
  correlators in four-dimensional SCFT}},  {\em Phys. Lett.} {\bf B472} (2000)
  323--331, [\href{http://xxx.lanl.gov/abs/hep-th/9910150}{{\tt
  hep-th/9910150}}].

\bibitem{Eden:2000gg}
B.~U. Eden, P.~S. Howe, E.~Sokatchev, and P.~C. West, {\it {Extremal and
  next-to-extremal n-point correlators in four- dimensional SCFT}},  {\em Phys.
  Lett.} {\bf B494} (2000) 141--147,
  [\href{http://xxx.lanl.gov/abs/hep-th/0004102}{{\tt hep-th/0004102}}].

\bibitem{Erdmenger:1999pz}
J.~Erdmenger and M.~Perez-Victoria, {\it {Non-renormalization of
  next-to-extremal correlators in N = 4 SYM and the AdS/CFT correspondence}},
  {\em Phys. Rev.} {\bf D62} (2000) 045008,
  [\href{http://xxx.lanl.gov/abs/hep-th/9912250}{{\tt hep-th/9912250}}].

\bibitem{D'Hoker:2000dm}
E.~D'Hoker, J.~Erdmenger, D.~Z. Freedman, and M.~Perez-Victoria, {\it
  {Near-extremal correlators and vanishing supergravity couplings in AdS/CFT}},
   {\em Nucl. Phys.} {\bf B589} (2000) 3--37,
  [\href{http://xxx.lanl.gov/abs/hep-th/0003218}{{\tt hep-th/0003218}}].

\bibitem{D'Hoker:2000vb}
E.~D'Hoker and B.~Pioline, {\it {Near-extremal correlators and generalized
  consistent truncation for AdS(4|7) x S(7|4)}},  {\em JHEP} {\bf 07} (2000)
  021, [\href{http://xxx.lanl.gov/abs/hep-th/0006103}{{\tt hep-th/0006103}}].

\bibitem{Aharony:2004xn}
O.~Aharony, A.~Giveon, and D.~Kutasov, {\it {LSZ in LST}},  {\em Nucl. Phys.}
  {\bf B691} (2004) 3--78, [\href{http://xxx.lanl.gov/abs/hep-th/0404016}{{\tt
  hep-th/0404016}}].

\bibitem{Lando:2003gx}
S.~K. Lando and A.~K. Zvonkin, {\it Graphs on surfaces and their applications},
  . Springer (2004), 403 p.

\bibitem{Dixon:1986qv}
L.~J. Dixon, D.~Friedan, E.~J. Martinec, and S.~H. Shenker, {\it The conformal
  field theory of orbifolds},  {\em Nucl. Phys.} {\bf B282} (1987) 13--73.

\bibitem{Lando2}
S.~K. Lando, {\it {Ramified coverings of the two-dimensional sphere and the
  intersection theory in spaces of meromorphic functions on algebraic curves}},
   {\em Russ. Math. Surv.} {\bf 57} (2002) 463--533.

\bibitem{Bouchard:2007hi}
V.~Bouchard and M.~Marino, {\it {Hurwitz numbers, matrix models and enumerative
  geometry}},  \href{http://xxx.lanl.gov/abs/0709.1458}{{\tt 0709.1458}}.

\end{thebibliography}\endgroup

\bibliographystyle{JHEP}
%\bibliographystyle{unsrt}
%\bibliography{ref}
%\bibliographystyle{apsrev}
%\bibliographystyle{plain}
%\bibliographystyle{utphys}

\end{document}